\documentclass[aps,twocolumn,groupedaddress,superscriptaddress]{revtex4-1}

\usepackage{graphicx,amsmath,amssymb}

\begin{document}

\begin{titlepage}
	
\title{Unified phase diagram of reversible-irreversible, jamming and yielding transitions in cyclically sheared soft sphere packings}
\author{Pallabi Das} 
\affiliation{Theoretical Sciences Unit, Jawaharlal Nehru Centre for Advanced Scientific Research, Bengaluru, India.}
\author{Vinutha H. A.}
\affiliation{Theoretical Sciences Unit, Jawaharlal Nehru Centre for Advanced Scientific Research, Bengaluru, India.}
\affiliation{Institute of Physics, Chinese Academy of Sciences, Beijing, China.}
\affiliation{Department of Chemistry, University of Cambridge, Cambridge, UK}
\author{Srikanth Sastry}
\affiliation{Theoretical Sciences Unit, Jawaharlal Nehru Centre for Advanced Scientific Research, Bengaluru, India.}

\begin{abstract}
Self-organization, and transitions from reversible to irreversible behaviour, of interacting particle assemblies driven by externally imposed stresses or deformation is of interest in comprehending diverse phenomena in soft matter. They have been investigated in a wide range of systems, such as colloidal suspensions, glasses, and granular matter. In different density and driving regimes, such behaviour is related to yielding of amorphous solids, jamming, and memory formation, \emph{etc.}  How these
phenomena are related to each other has not, however, been much studied.  In
order to obtain a unified view of the different regimes of behaviour,
and transitions between them, we investigate computationally the
response of soft sphere assemblies to athermal cyclic shear
deformation over a wide range of densities and amplitudes of shear
deformation. Cyclic shear deformation induces 
transitions from reversible to irreversible behaviour in both unjammed and jammed soft sphere packings. Well
above isotropic jamming density ($\bf{\phi_J}$), this transition
corresponds to yielding.  In the vicinity of the jamming point, up to
a higher density limit we designate ${\bf \phi_J^{cyc}}$, an unjammed phase
emerges between a localised, \emph{absorbing} phase, and a diffusive,
{\emph irreversible} phase.  The emergence of the unjammed phase signals
the shifting of the jamming point to higher densities as a result of
annealing, and opens a window where shear jamming becomes possible for
frictionless packings. Below $\bf{\phi_J}$, two distinct localised states,
termed point and loop reversibile, are observed.  We characterise in
detail the different regimes and transitions between them, and obtain
a unified density-shear amplitude phase diagram.
\end{abstract}

\maketitle
\end{titlepage}

\section{Introduction}
The response of disordered assemblies of particles to externally imposed stresses or deformation is of importance in wide ranging investigations addressing transitions between rigid and flowing states of soft matter systems, and characterizing the rigid and flowing states. Questions in this regard pertain to the rheology of complex fluids, in particular, dense particulate suspensions\cite{mason1996,brown2012role,peters2016direct,mari2014shear,wyart2014discontinuous}, jamming in granular matter\cite{liu2010jamming,bi2011,ren2013reynolds,coulais2014,baity2016,behringer2018physics,vinutha2016geometric,vinutha2019force}
and the mechanical behaviour, yielding and shear banding in amorphous solids (from metallic glasses to yield stress soft materials)\cite{falk2011deformation,karmakar2010statistical,urbani2017shear,jin2018,jaiswal2016mechanical,pk2017,parisi2017shear,divoux2016shear,bonn2017yield,vasisht2017emergence,kawa2016}. Investigations on these questions inform geophysical phenomena such as earthquakes and landslides \cite{jagla2014viscoelastic,uhl2015universal,jerolmack2019viewing}, material properties of metallic glasses\cite{sun2016thermomechanical,priezjev2019effect}, the control of rheological response of suspensions \cite{ness2018shaken}, origins of irreversible behaviour\cite{pine2005,corte2008,schreck2013particle,regev2013}, and memory formation in a variety of condensed matter systems\cite{keim2018memory,royer2015precisely,fiocco2014encoding,adhikari2018memory,lavrentovich2017period,paulsen2014multiple,keim2011}, to name a few examples. A number of experimental and computational investigations addressing such questions have employed the protocol of oscillatory or cyclic shear deformation, often in the {\it athermal} limit, when thermal fluctuations do not play a significant role. 

At low densities, cyclically sheared colloidal suspensions (and models thereof) exhibit a continuous transition from reversible to irreversible behaviour \cite{pine2005,corte2008}, with time scales to reach steady states apparently diverging at the transition. At high densities, dense glasses exhibit a sharp, discontinuous yielding transition  \cite{fiocco2013oscillatory,pk2017,regev2013,priezjev2013,regev2015,kawa2016}, nevertheless with apparent divergence of times to reach the steady state. These two regimes also display memory effects with \cite{keim2011,paulsen2014multiple,fiocco2014encoding,adhikari2018memory} significant differences in behaviour. At intermediate densities, where jamming behaviour has been explored \cite{ohern-2003,pinaki-2010,ozawa2012jamming}, mechanical measurements reveal interesting phenomena such as softening and yielding \cite{coulais2014,dagois2017softening}, which are not fully or well characterized. Thus, depending on the density regime, one finds transitions from reversible to irreversible behaviour with varying characteristics.

Given the diversity of observed behaviours, it is of interest to comprehend  the relationship between these seemingly different but related phenomena. Such a comprehensive understanding is hampered by the fact that these different regimes have been probed on physically very different experimental systems (colloidal suspensions, granular matter, molecular or atomic glasses), or different computational models and methods. It is thus desirable to interrogate all these phenomena in a single system to establish the relationship between different regimes of behaviour. In the present work, we address this goal by studying the model system of athermal soft sphere assemblies, which have been used to model, depending on the density of the packings, the behaviour of colloidal suspensions, granular matter and dense glasses. We study the behaviour of a binary (50:50) mixture of soft spheres interacting through a harmonic potential, subjected to athermal cyclic deformation of varying amplitudes ( see Appendix for details) over a density range that encompasses all these regimes. By doing so, we observe reversible to irreversible transitions at the high and low density regimes, analogous to previous studies, but our results illuminate several new aspects not previously addressed. At intermediate densities, we further observe a window in which cyclic shear deformation unjams initially jammed configurations. This window is sandwiched between a reversible regime at low amplitudes and an irreversible regime at high amplitudes. Interestingly, this unjamming window allows us to probe a phenomenon that hitherto has largely been addressed in the context of frictional granular packings, namely shear jamming. Owing to the unjamming we obtain by cyclic shear deformation, we can study shear jamming in a frictionless system. We present our findings below by considering three distinct density regimes, and finally, construct a unified phase diagram that integrates all the observed regimes and transitions among them. Related results in two dimensions are presented in \cite{Nagasawa2019}.

\begin{figure*}[ht!]
\centering
\includegraphics[width=.32\linewidth]{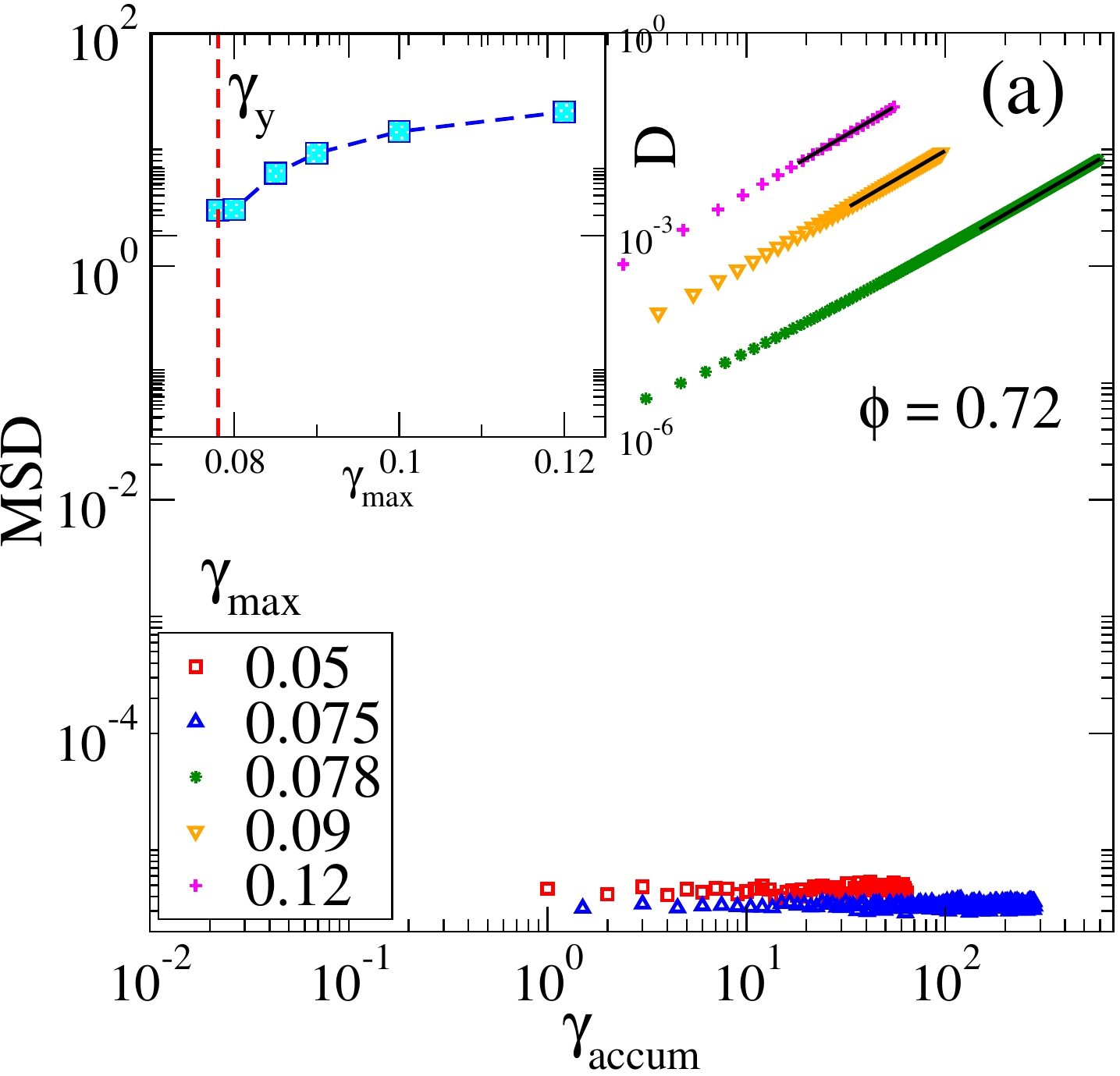}
\hspace{0.5cm}
\includegraphics[width=.34\linewidth]{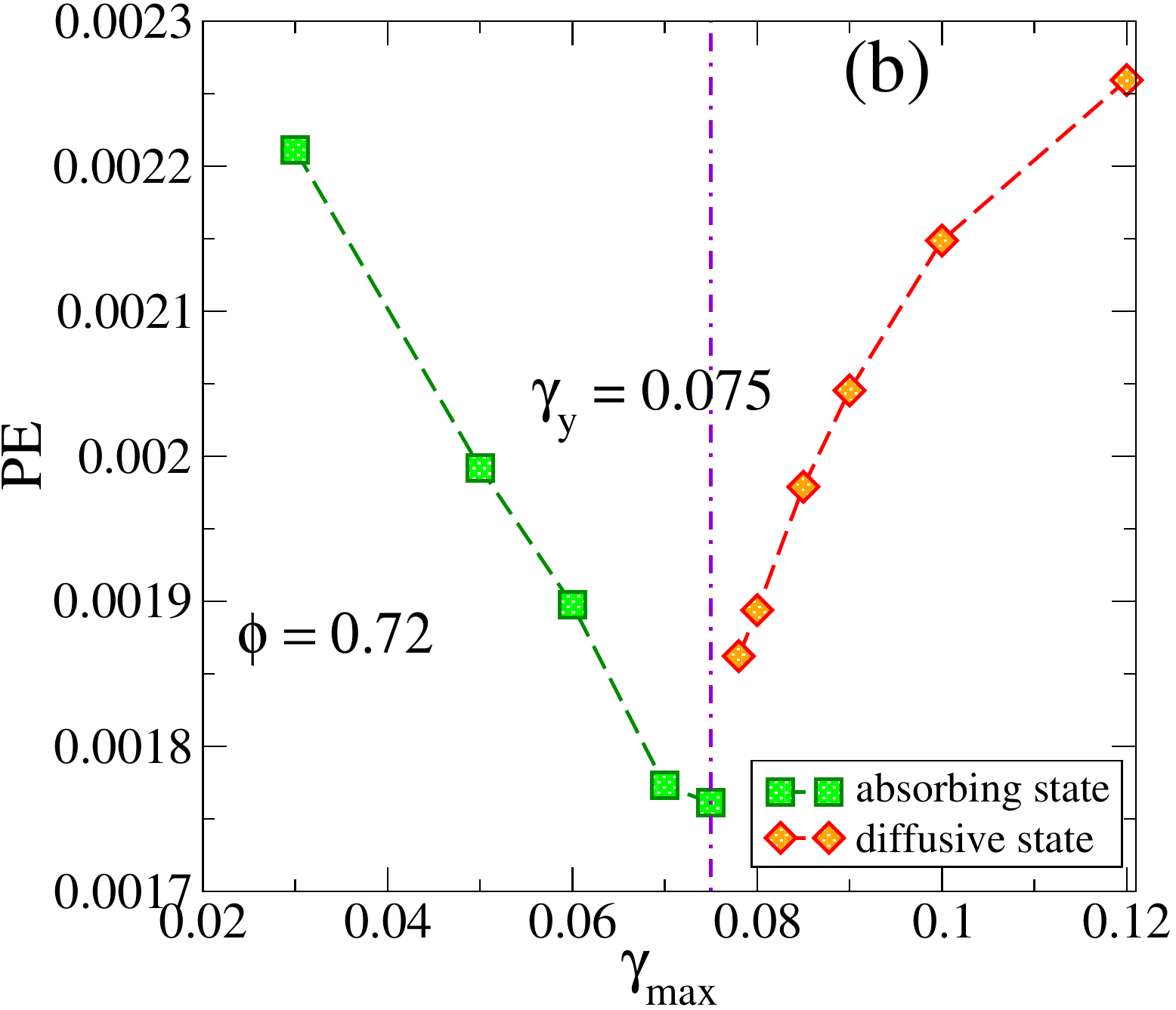}
\caption{ {\bf (a)} Mean square displacement (MSD) as a function of $\gamma_{accum}$, shown for different strain amplitudes. MSD shows a diffusive behavior above $\gamma_{max}=0.075$. (Inset {\bf (a)}): Above $\gamma_{max}=0.075$, the diffusion coefficient (D) jumps from zero to a finite value. {\bf (b)} The steady state potential energy (PE) value as a function of $\gamma_{max}$ is shown. The potential energy attains a minimum value at the yielding strain amplitude, indicated as $\gamma_y = 0.075$. At the yielding strain, the potential energy also shows a discontinuous jump.} 
\label{msd072}
\end{figure*}

\section{Reversible-irreversible and yielding transition}
    The soft sphere model system well above $\phi_J$ has been studied and observed to behave as a good glass former \cite{berthier2009compressing,berthier2009glass}. It has been shown previously that amorphous solids under cyclic shear deformation yield at a well defined strain at which they undergo (in the steady states reached after repeated cycles of strain, which we always focus on unless otherwise stated) a transition from a reversible state (where particles return to the same position after each cycle) to an irreversible, diffusive, state \cite{fiocco2013oscillatory,regev2013,pk2017}. Therefore we identify the yield strain at which the mean squared displacement (MSD) of particle positions measured stroboscopically changes from zero (absorbing phase, A) to finite values (diffusive or yielding phase Y) in a discontinuous fashion. The yield strain can also be identified by the non-monotonic, and discontinuous change in the steady state value of the potential energy (PE) \cite{regev2013,pk2017,parmar2019strain}. In Fig. \ref{msd072}(a), we show the mean squared displacement (MSD) for different shear amplitudes at one high packing fraction, $\phi = 0.72$, above the jamming density $\phi_J$ (the {\it minimum} value of $\phi_J$ in this system is estimated to be $0.648$). For small amplitudes, the MSD is negligible and non-increasing with cycles, whereas above a strain amplitude of  $\gamma_{max} = 0.075$, it exhibits a linear increase with accumulated strain $\gamma_{acc} = 4 \times \gamma_{max} \times N_{cycles}$ (Fig. \ref{msd072}(a)). The diffusion coefficients obtained from $MSD = 6 D \gamma_{acc}$, shown in the inset of Fig. \ref{msd072}(a), display  a discontinuous jump at $\gamma_{max} = 0.075$, as observed in previous studies \cite{kawa2016,parmar2019strain}.  In Fig. \ref{msd072}(b), we show the steady state per particle potential energy (PE) as a function of the strain amplitude, which shows a discontinuity at $\gamma_{max} = 0.075$, as seen previously for a binary Lennard-Jones glass \cite{pk2017,parmar2019strain}.  We identify the steady state by monitoring PE as a function of $\gamma_{accum}$ (see Appendix section $3$). From these two characterizations, we identify $\gamma_{y} = 0.075$ as the yield strain value. The same behaviour is observed over a range of densities (data not shown) at roughly a constant $\gamma_{max} \approx 0.075$ and is consistent with the characterization of the yielding transition in previous work. 


\section{Unjamming and shear jamming above the isotropic jamming density}

We next consider the density range just above the isotropic jamming density, $\phi_J$, estimated to be $0.648$ using previously employed methods \cite{ohern-2003,pinaki-2010} of compressing low density configurations. As noted before, the jamming density is not unique \cite{pinaki-2010,ozawa2012jamming,kumar2016memory,jin2018}, and jammed packings just above $\phi_J$ exhibit peculiar mechanical properties \cite{ohern-2003,wyart2005g}. We thus expect interesting responses to cyclic deformation in this regime. 

\begin{figure*}[ht!]
\centering
\includegraphics[width=.32\linewidth]{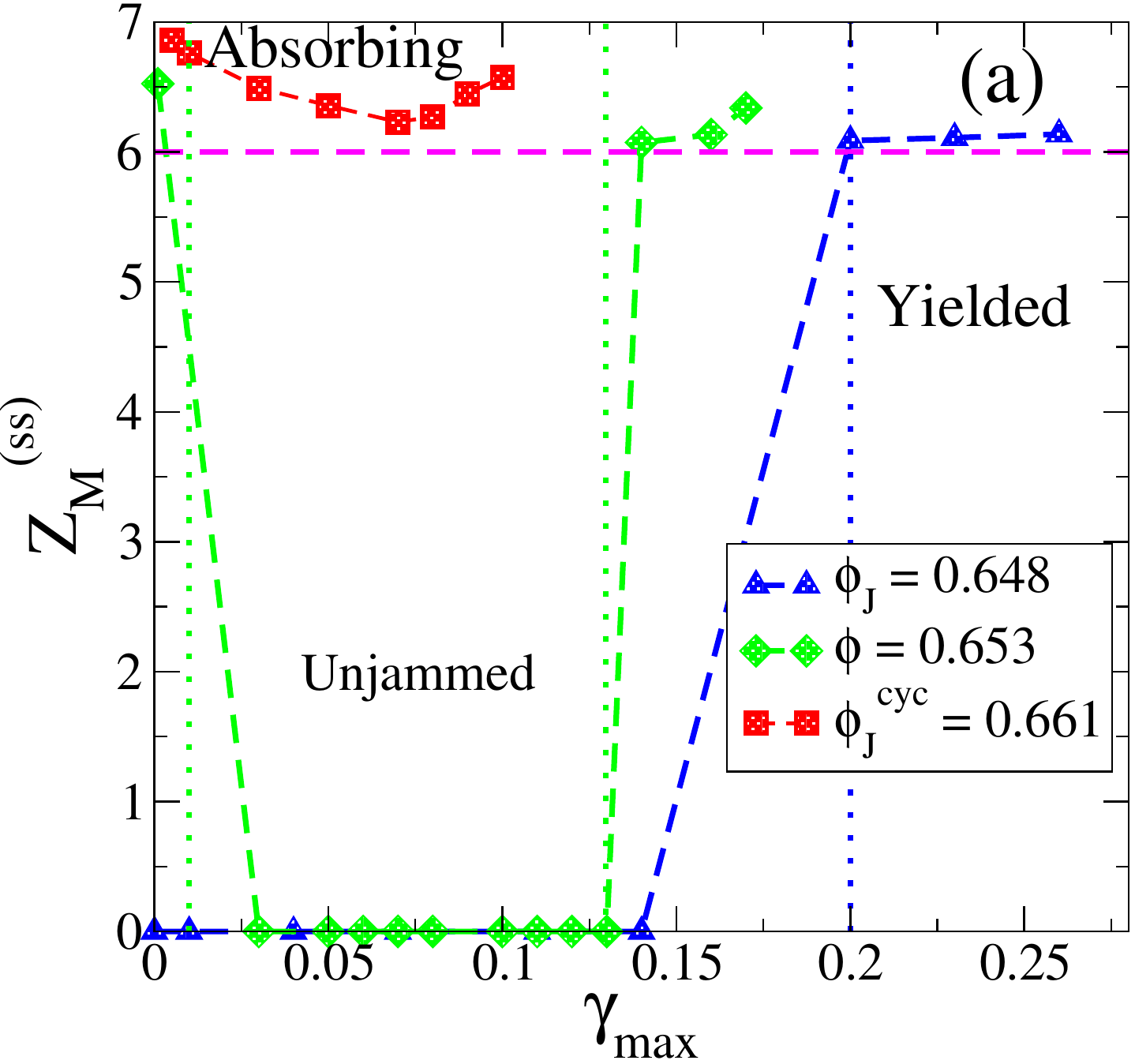}
\includegraphics[width=.34\linewidth]{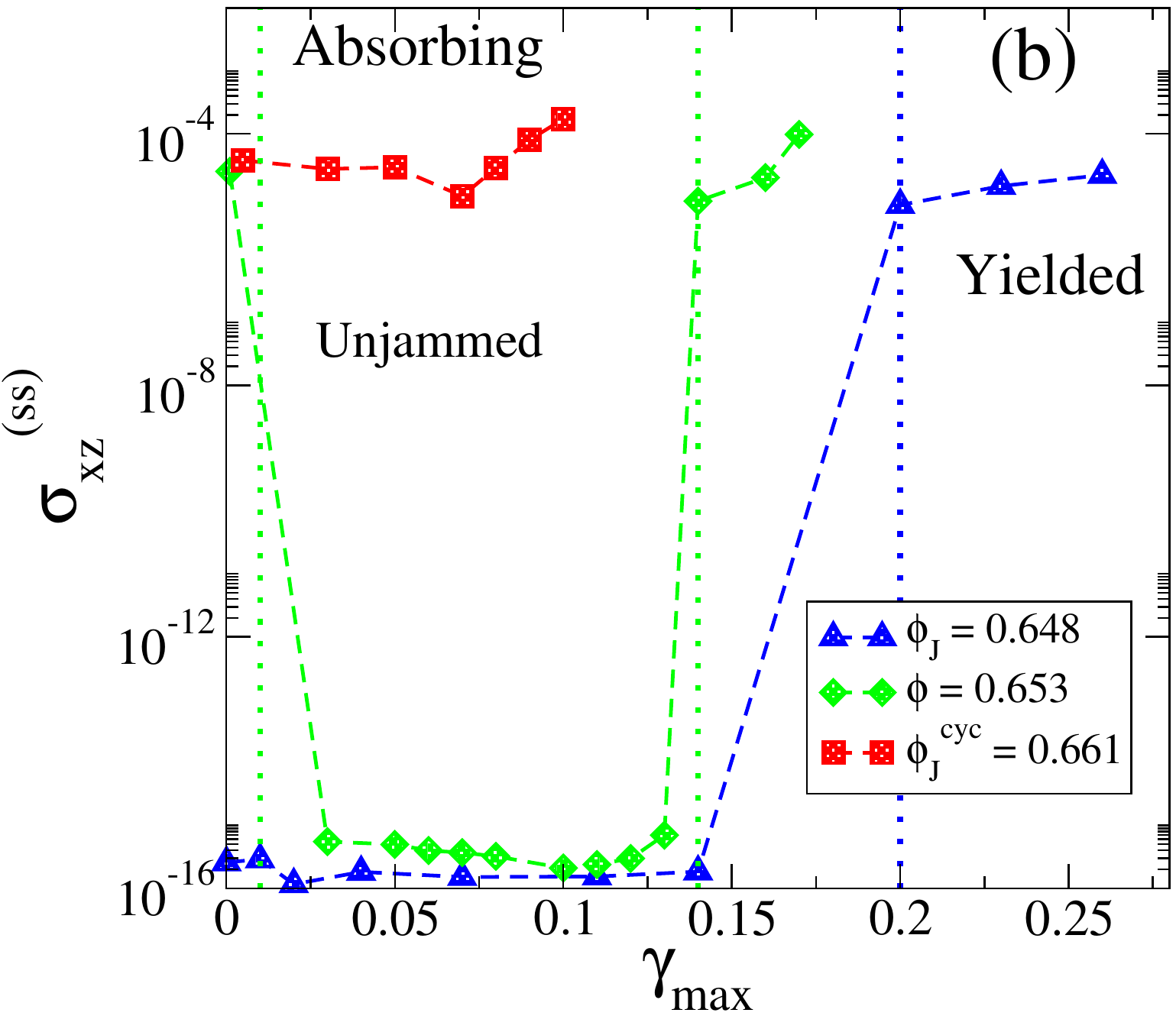}
\caption{The average mechanical contact number in the steady state, computed with only force bearing contacts, jumps to $0$ in the unjamming regime ({\bf (a)}), as does the the steady state shear stress $(\sigma_{xz})_{ss}$ ({\bf(b)}).} 
\label{zunjmall} 
\end{figure*}

We observe that for a range of packing fractions close to $\phi_J$ the system unjams for an intermediate range of amplitudes, {\it i.e.,} shear stress goes to zero. In Fig. \ref{zunjmall} (a), we show the steady state value of the stroboscopic average mechanical contact number $Z_M$ (defined in Appendix, section $1$), and the shear stress in Fig. \ref{zunjmall}(b), for different densities, as a function of strain amplitude. The corresponding behaviour of the geometric contact number $Z$ and the non-rattler contact number ($Z_{NR}$) (see Appendix, section 1) is shown in the Appendix, along with the dependence of $Z$ on the minimization protocol and the  variation of $Z_{NR}$ with $\gamma_{accum}$ (Appendix, sections $4$ and $5$). The shear stress jumps to zero when $Z_M$ (also $Z$, $Z_{NR}$) goes below $2D(=6)$, the {\it isostatic} value for jamming in frictionless packings. The unjamming range of shear amplitudes is largest at  $\phi_J$, and decreases with increasing density, till it vanishes at $\phi = 0.661 $. At $\phi \geq \phi_J$, both for high amplitudes and very low amplitudes, the system has a finite value of stress. We identify the finite stress packings at low amplitudes as the absorbing phase, and at high amplitudes as the yielded phase. We interpret the presence of an unjamming regime as a reflection of the jamming density moving to higher values upon cyclic deformation.  Thus, we identify $\phi=0.661$ as the $\phi_J^{cyc}$ or the cyclic jamming point, the highest jamming density obtained under athermal cyclic shear. Noting that $\phi_J^{cyc}$ equals the highest density obtained in \cite{pinaki-2010}, it is tempting to treat it as the highest possible jamming density, but  further exploration of possible protocols is required to draw such a conclusion. Our results concerning unjamming are consistent with the observation of softening behaviour of jammed solids close to $\phi_J$, for an intermediate range of strain amplitudes \cite{dagois2017softening}.

\begin{figure}[ht!]
\centering
\includegraphics[width=.48\linewidth]{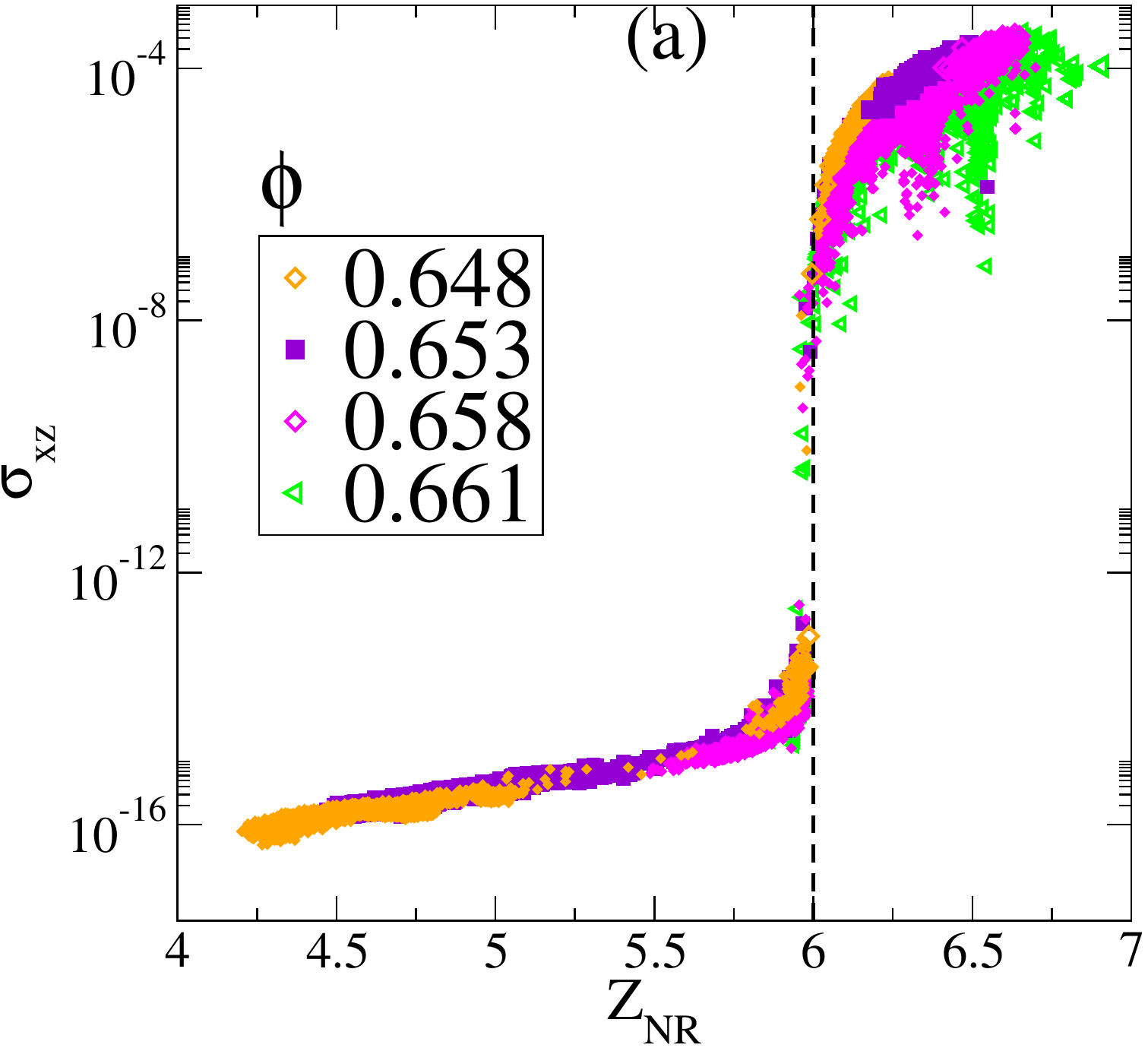}
\includegraphics[width=.48\linewidth]{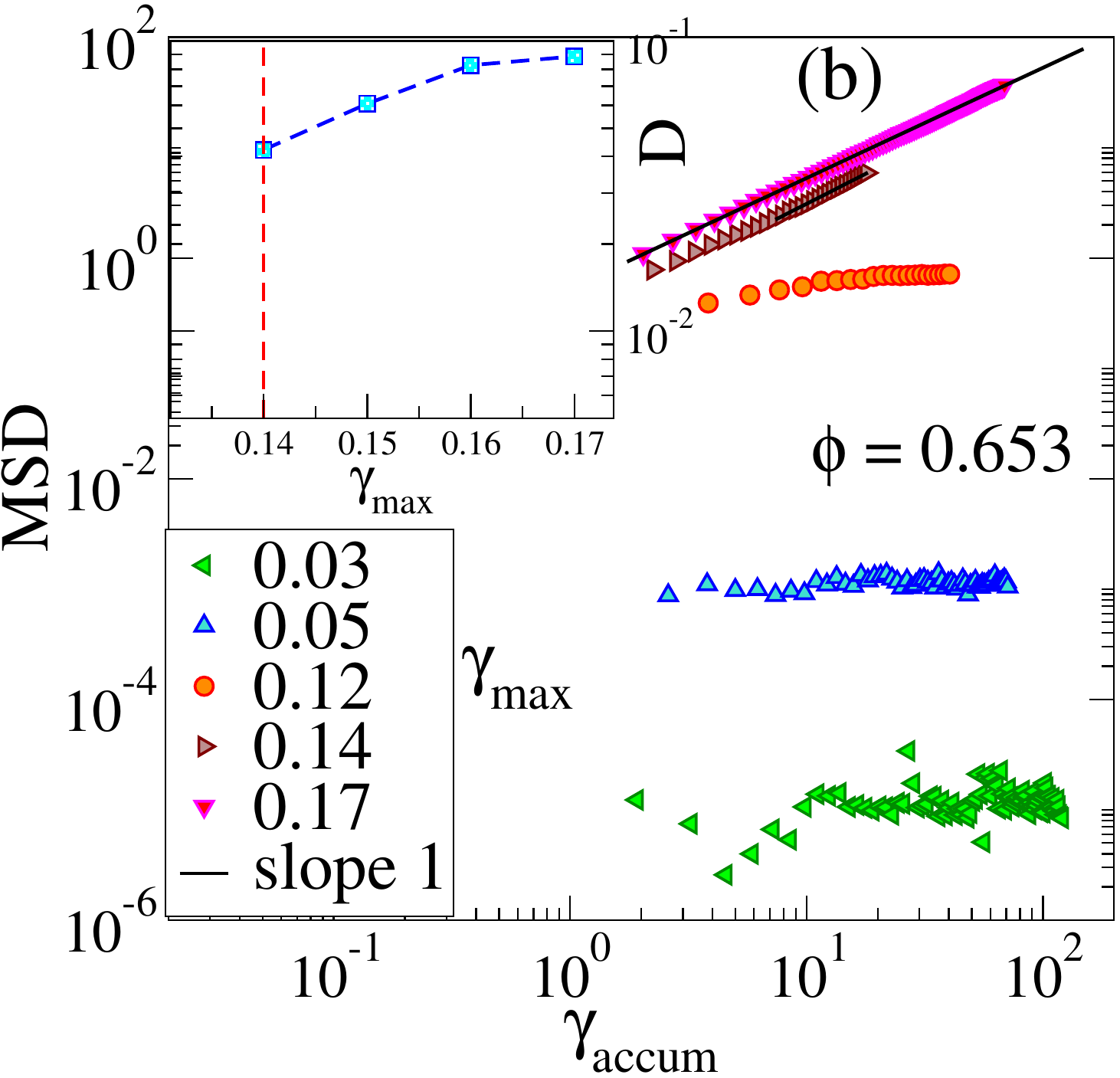}
\caption{(a) Shear stress ($\sigma_{xy}$) \emph{vs} the average contact number ($Z_{NR}$), for stroboscopic configurations, shown for different densities and strain amplitudes. Packings in the range $ \phi_J \geq \phi < \phi_J^{cyc}$ exhibit two distinct branches of finite or zero stress. The shear stress jumps to a finite value at $Z_{NR} = 6$ (b) The MSD for the jammed (absorbing) and unjammed regimes exhibit non-diffusive behaviour whereas the shear jamming regime exhibits diffusive behaviour. The diffusion coefficients (inset of (b)) show a discontinuity at the re-entrant shear jamming transition.}
\label{sigz} 
\end{figure}

In Fig. \ref{sigz} (a), we show the shear stress as a function of $Z_{NR}$, for one sample per density and strain amplitude (in the same range as Fig. \ref{zunjmall}). Each point in the scatter plot corresponds to all the stroboscopic configurations, from the initial configuration up to cycle numbers when the steady state is reached. For $\phi < \phi_J^{cyc}$,  Fig. \ref{sigz} (a) exhibits two branches, one with close to zero stress and one with finite stress when $Z_{NR} \geq 6$, above the isostatic contact number. The re-entrant finite stress regime with $Z_{NR} \geq 6$ at high strain amplitudes corresponds to shear jamming, explored in great detail for frictional granular packings \cite{zhang2010,bi2011,vinutha2016geometric,vinutha2019force,otsuki2018shear,ishima2019dilatancy,mobius2014}, but also as a possibility in recent times for frictionless packings \cite{kumar2016memory,baity2016,jin2018}. Our results open a novel route to generating unjammed packings above the isotropic jamming point $\phi_J$, which can shear jam upon the application of shear. 

Since the shear jamming is analogous to and contiguous with the yielding transition at higher densities, it is interesting to analyse their comparison further. To this end, we show in Fig \ref{sigz}(b) the MSD in the steady state. for a range of shear amplitudes spanning the absorbing, unjamming and shear jamming regimes. 
The MSD shown correspond to zero diffusivity for the absorbing (jammed) and the unjamming regimes (and therefore does not distinguish these phases), but a finite diffusivity in the shear jamming phase. As shown in the inset of Fig \ref{sigz}, the diffusivities display a discontinuous jump upon entering the shear jamming regime, analogous to the irreversible, yielded regime at higher densities (see Fig. \ref{msd072}), above $\phi_J^{cyc}$. Although diffusivity does not distinguish the absorbing and unjammed regimes, they are distinguished by the presence of zero {\it vs.} finite stresses.



\section{Reversible-irreversible transitions below the isotropic jamming density}

We next study cyclically sheared sphere assemblies below $\phi_J$, which show more complex behaviour in the reversible regime \cite{schreck2013particle} than initially analysed for colloidal suspensions \cite{pine2005,corte2008}. Schreck {\it et al.} \cite{schreck2013particle} showed that for such packings, two kinds of reversible states are present, referred to as point reversible states (PR) and loop reversible states (LR). In PR states, particles self organize during the initial cycles of strain so that they do not collide with each other. In the LR state, particle continue to collide in the steady state, but return to their original positions at the end of each cycle. For packings below $\phi_J$, potential energies and stresses are always zero. Thus, we characterise the trajectories of particles in the steady state in various ways to distinguish different phases and transitions. In addition the MSD, we compute the non-affine path length ($L$) (the length of the path traversed by particles in excess of of affine displacements due to global shear strain), and the percentage of new collisions ($C_{new}$) in a given cycle, compared to a reference cycle in the steady state (see Appendix, section $1$). As shown in Fig. \ref{point0627}(a), the MSD saturates for small amplitudes and shows diffusive behaviour at higher amplitudes, identified as the irreversible phase (IR). The diffusion coefficient shows a discontinuity (inset of Fig. \ref{point0627}) in a manner analogous to the shear jamming and yielding transitions. MSD fails to identify the transition from the point reversible to the loop reversible state. However, the non-affine path length $L$ does: $L$ is negligible for point reversible states whereas it is finite for loop reversible states. The number of new collisions $C_{new}$ further distinguishes the reversible states from the irreversible states, being zero for reversible states and finite for irreversible states. As shown in Fig. \ref{point0627} (b), the combination of $L$ and $C_{new}$ helps organize all the regimes we observe across densities into three groups: (i) vanishing $L$, $C_{new}$ (point reversible), (ii) finite $L$, vanishing $C_{new}$ (loop reversible, unjamming and absorbing states), and (iii) finite $L$ and finite $C_{new}$ (irreversible, shear jammed and yielded states). Additional characterization given in Appendix, section $6$, shows that $L$ changes discontinuously both at the PR-LR, and LR-IR transitions, and $C_{new}$ jumps at the LR-IR transition. Surprisingly, the number of collisions, and the fraction of  particles undergoing collisions, change discontinuously at the PR-LR transition, at variance with expectations based on previous work \cite{pine2005,corte2008}. The time taken to reach steady state (measured through the decay of $L$ (Appendix section 7)), shows non-monotonic change across both the PR-LR and LR-IR transitions, consistently with expectations \cite{pine2005,corte2008,fiocco2013oscillatory,pk2017} for reversible-irreversible transitions. 

\begin{figure}[ht!]
\centering
\includegraphics[width=0.46\linewidth]{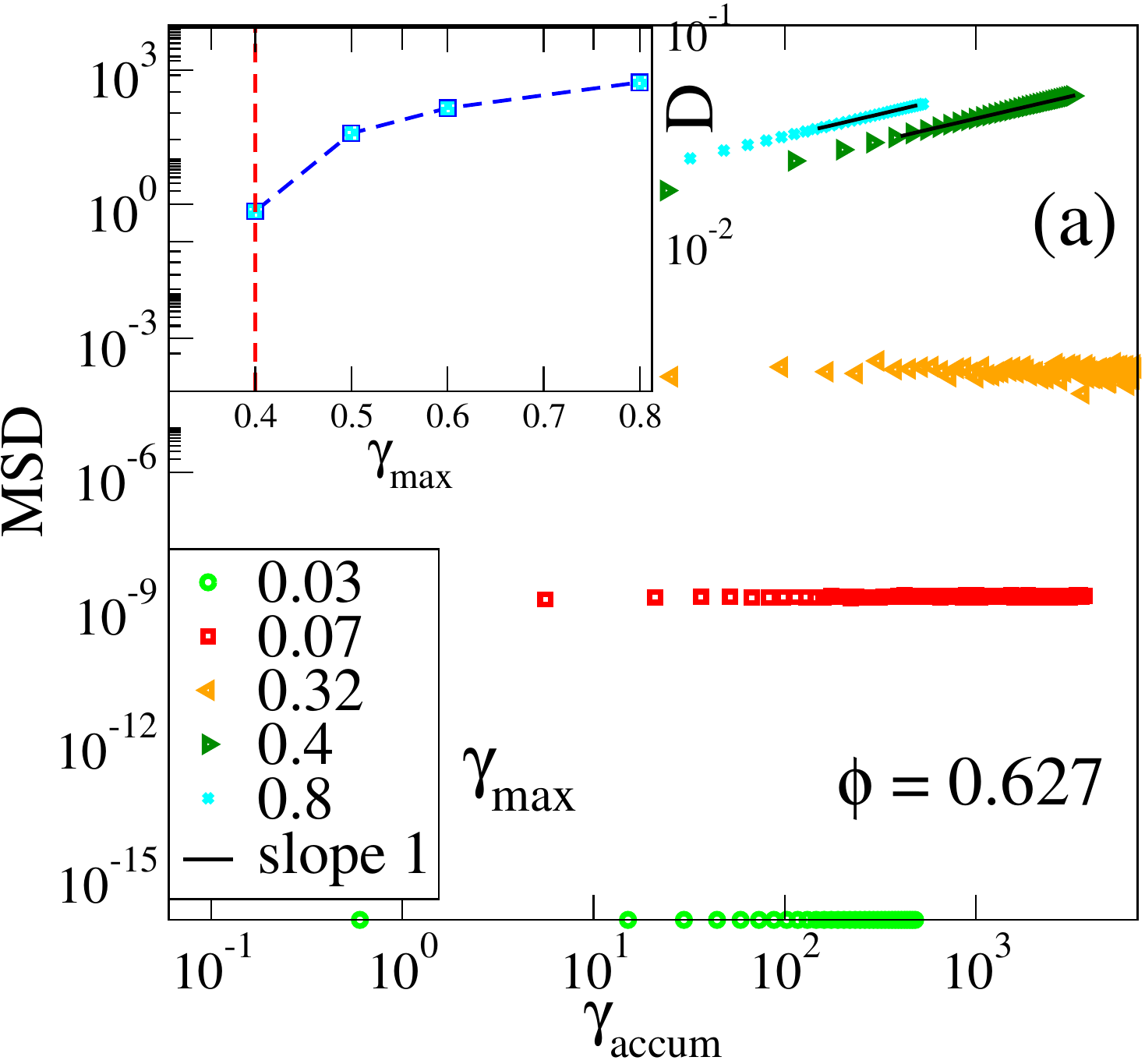}
\includegraphics[width=.46\linewidth]{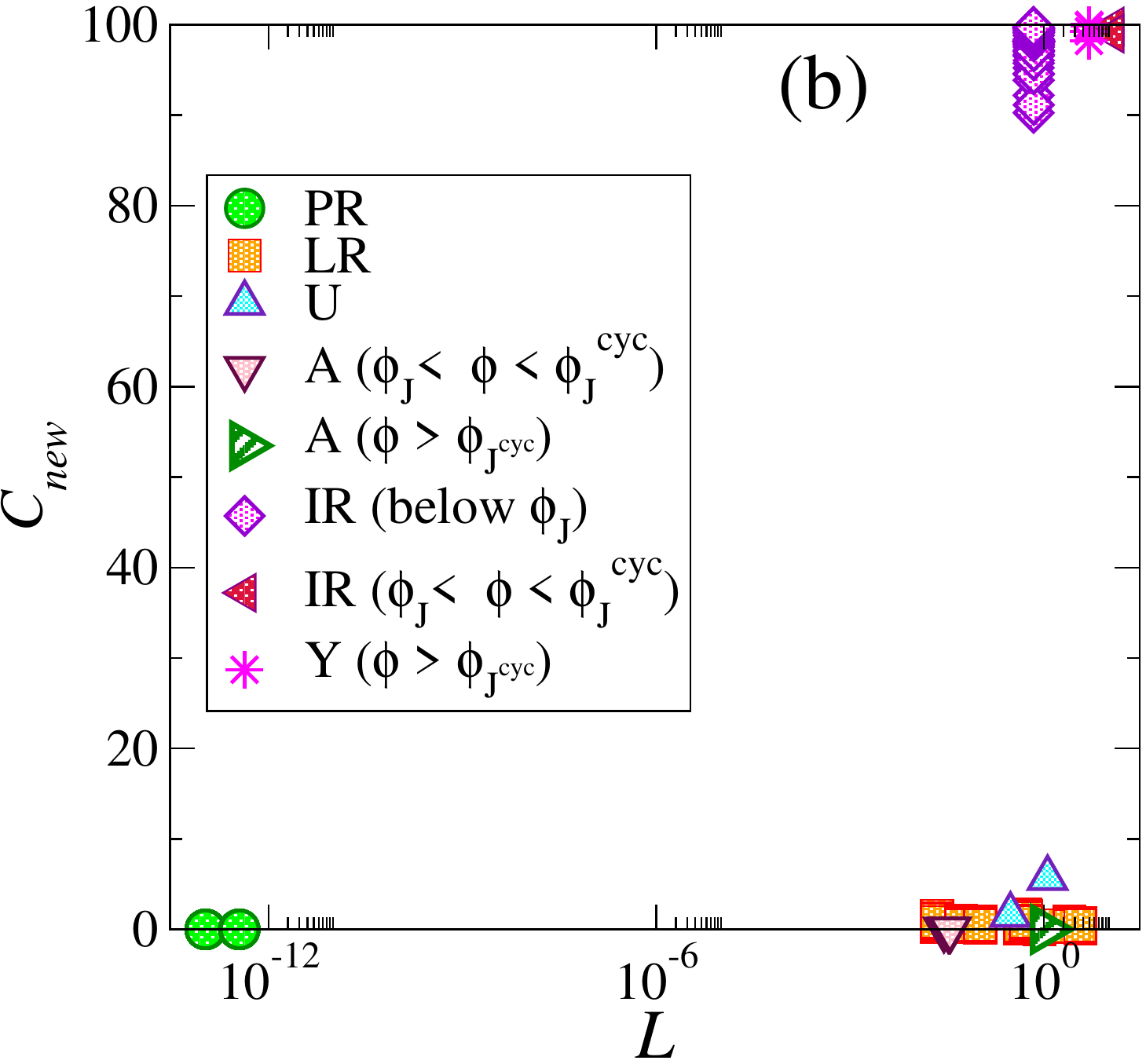}
\caption{(a) MSD as a function of accumulated strain, shown at $\phi=0.627$ for different amplitudes (a). We show a discontinuous jump in diffusivity  as a function of strain amplitude. (b) Plot of the non-affine path length $L$ and the percentage of new collisions $C_{new}$, which clearly differentiate all the different phases across $\phi_J$. PR: Point reversible; LR: Loop reversible; IR: Irreversible; Y: Yielded phase; U: Unjammed phase; A: Absorbing phase.} 
\label{point0627}
\end{figure}

  

The reversible-irreversible transition line for packings above $\phi_J$ corresponds to the yielding and shear jamming transitions, whereas below $\phi_J$ no such obvious correspondence can be drawn. There are recent studies suggesting the possibility of connecting the reversible-irreversible transition line to the shear jamming transition in frictional packings \cite{vinu2016,vinutha2019force,zhang2010,otsuki2018shear}, but further analysis is needed make such a connection firmly. 


\section{Phase diagram}

\begin{figure}[htp]
\centering
\includegraphics[width=.46\linewidth]{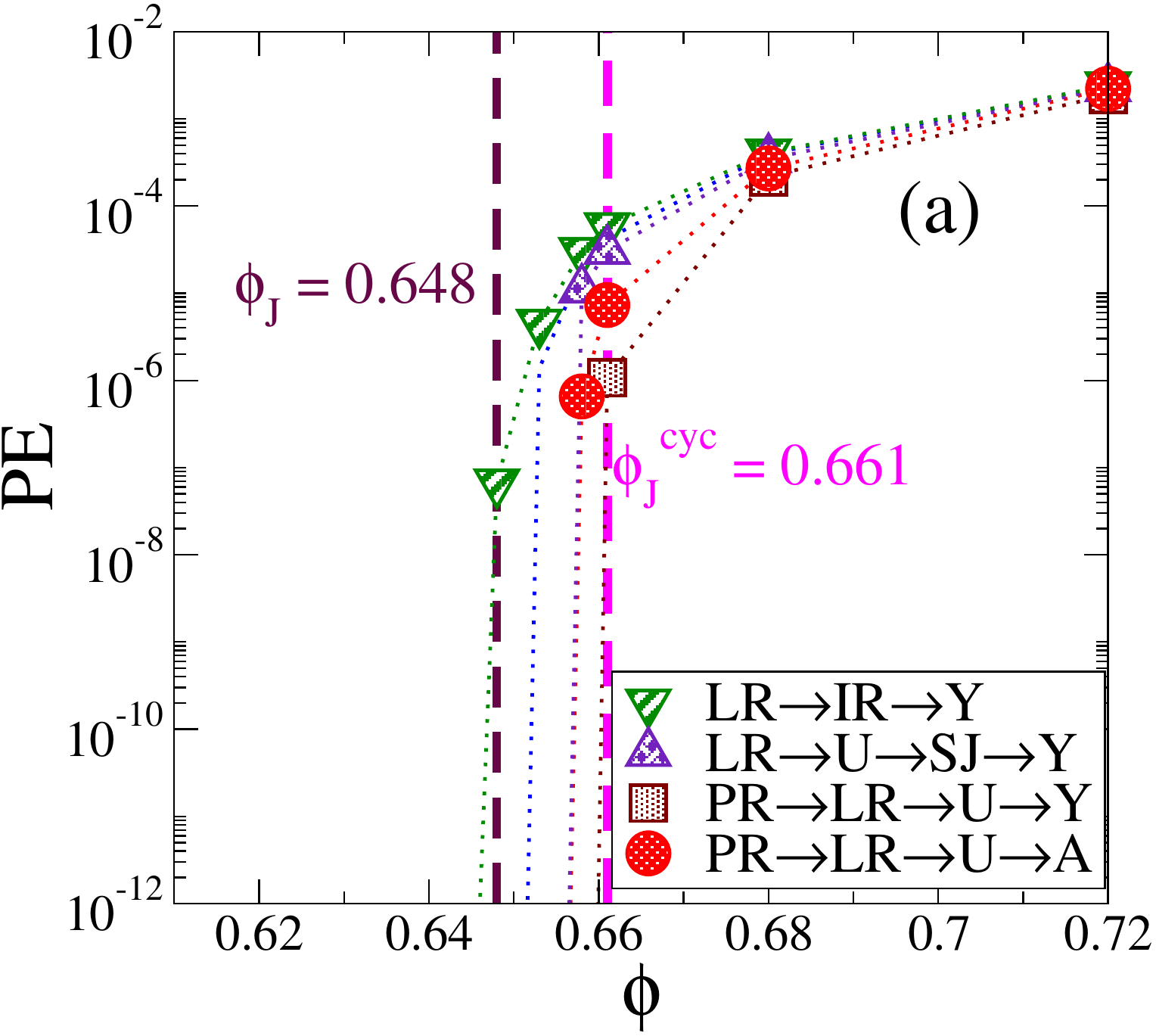}
\includegraphics[width=.47\linewidth]{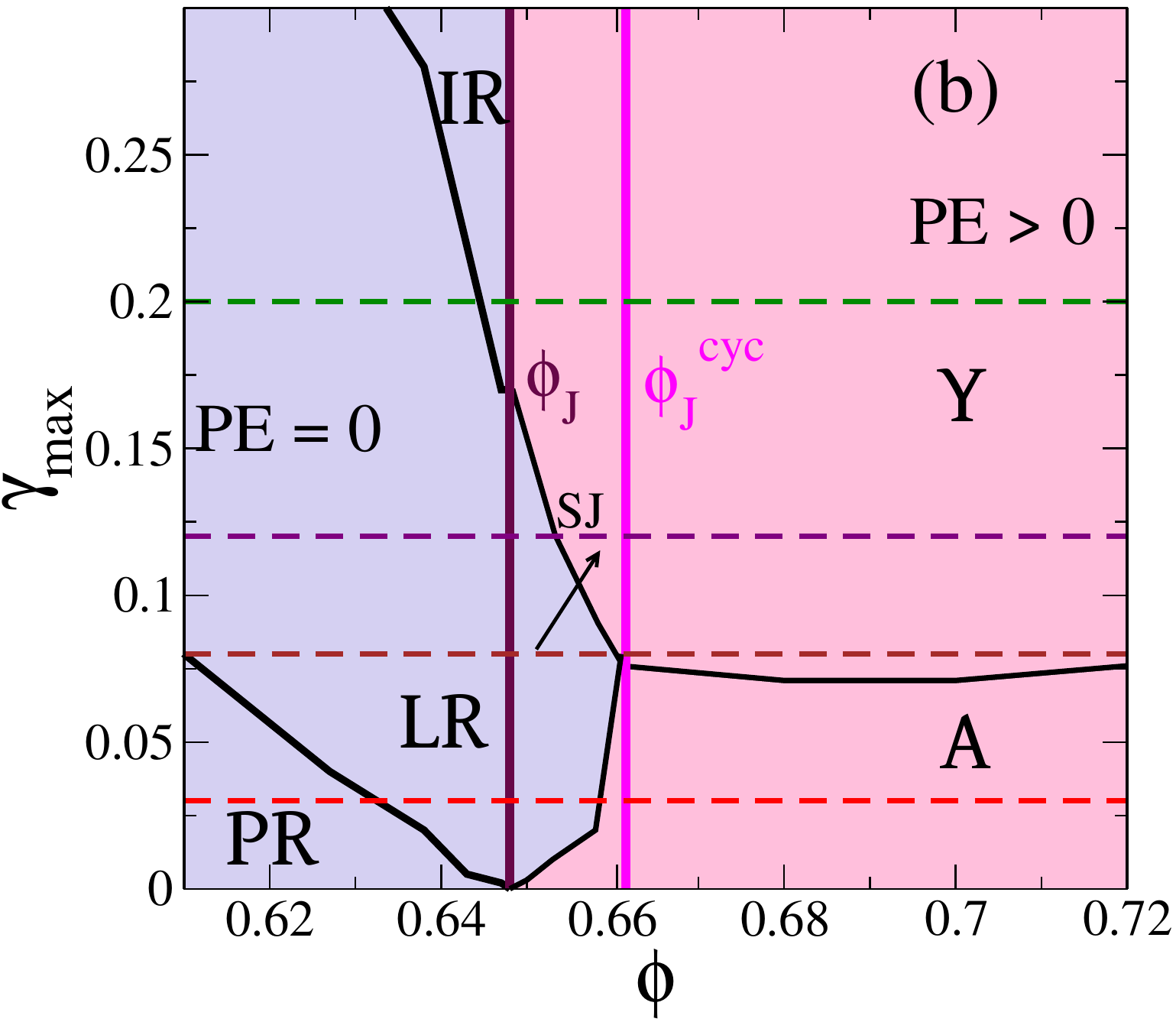}
\caption{Average stroboscopic steady state value of the potential energies shown as a function of $\phi$, for four different $\gamma_{max}$ (a). At these $\gamma_{max}$ values, potential energy shows a discontinuous change to finite values, but at different densities depending on  $\gamma_{max}$. For small  $\gamma_{max}$, the jump occurs at the transition from the unjamming to absorbing or shear jamming states, whereas at larger  $\gamma_{max}$, it occurs at $\phi_{J}$ (b). } 
\label{pephi}
\end{figure}

\begin{figure*}
\centering
\includegraphics[width=.24\linewidth]{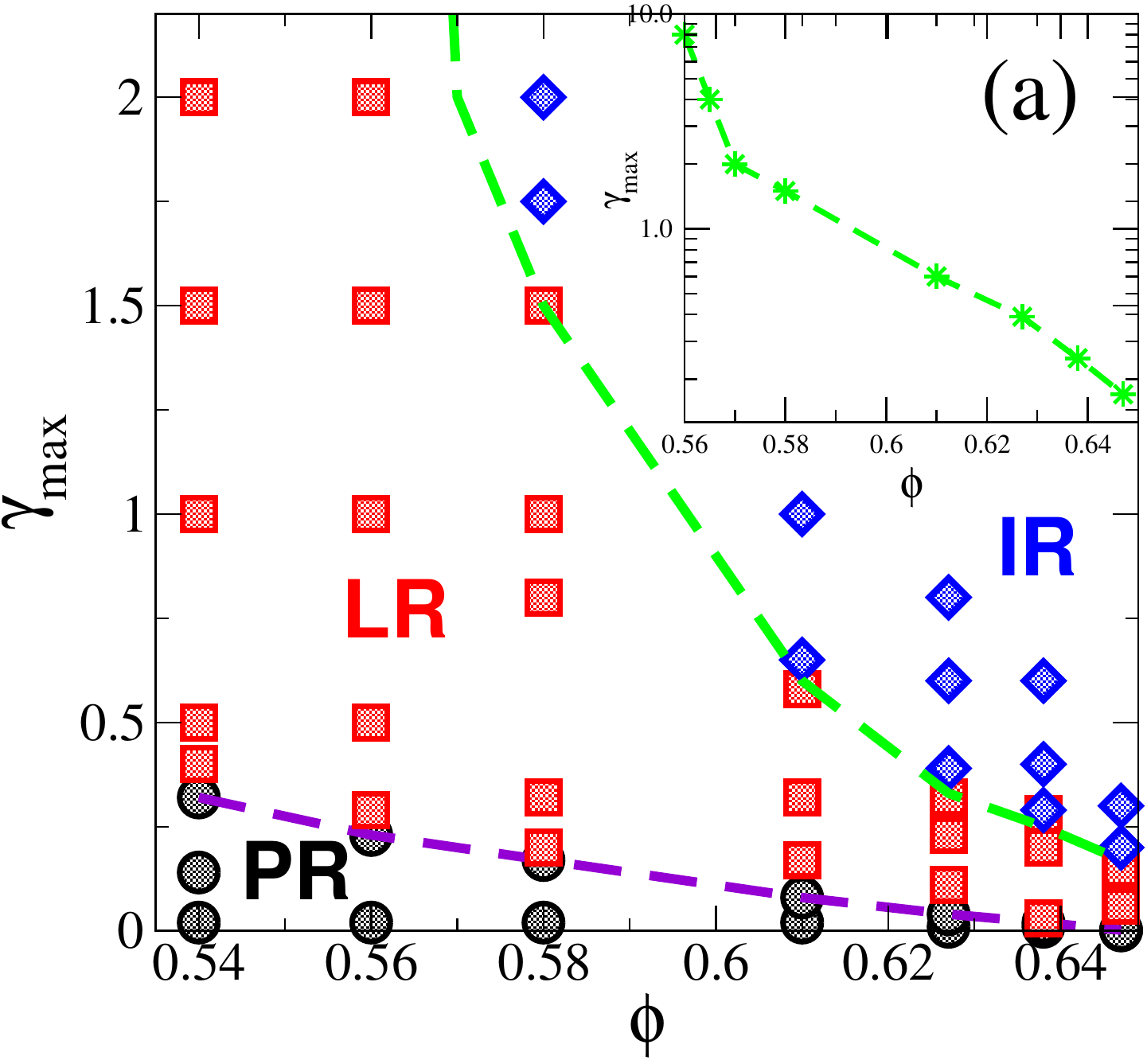}
\includegraphics[width=.24\linewidth]{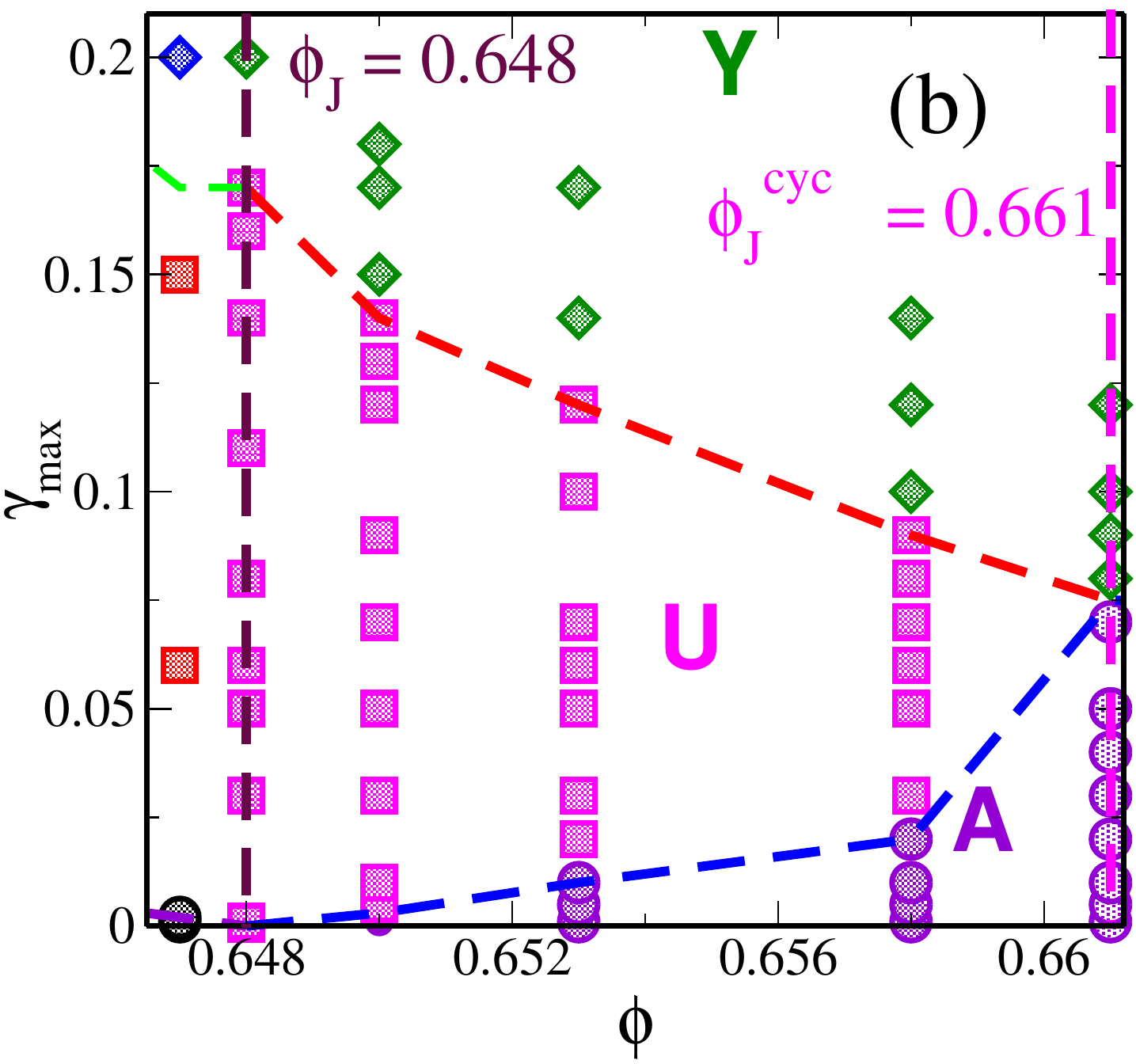}
\includegraphics[width=.24\linewidth]{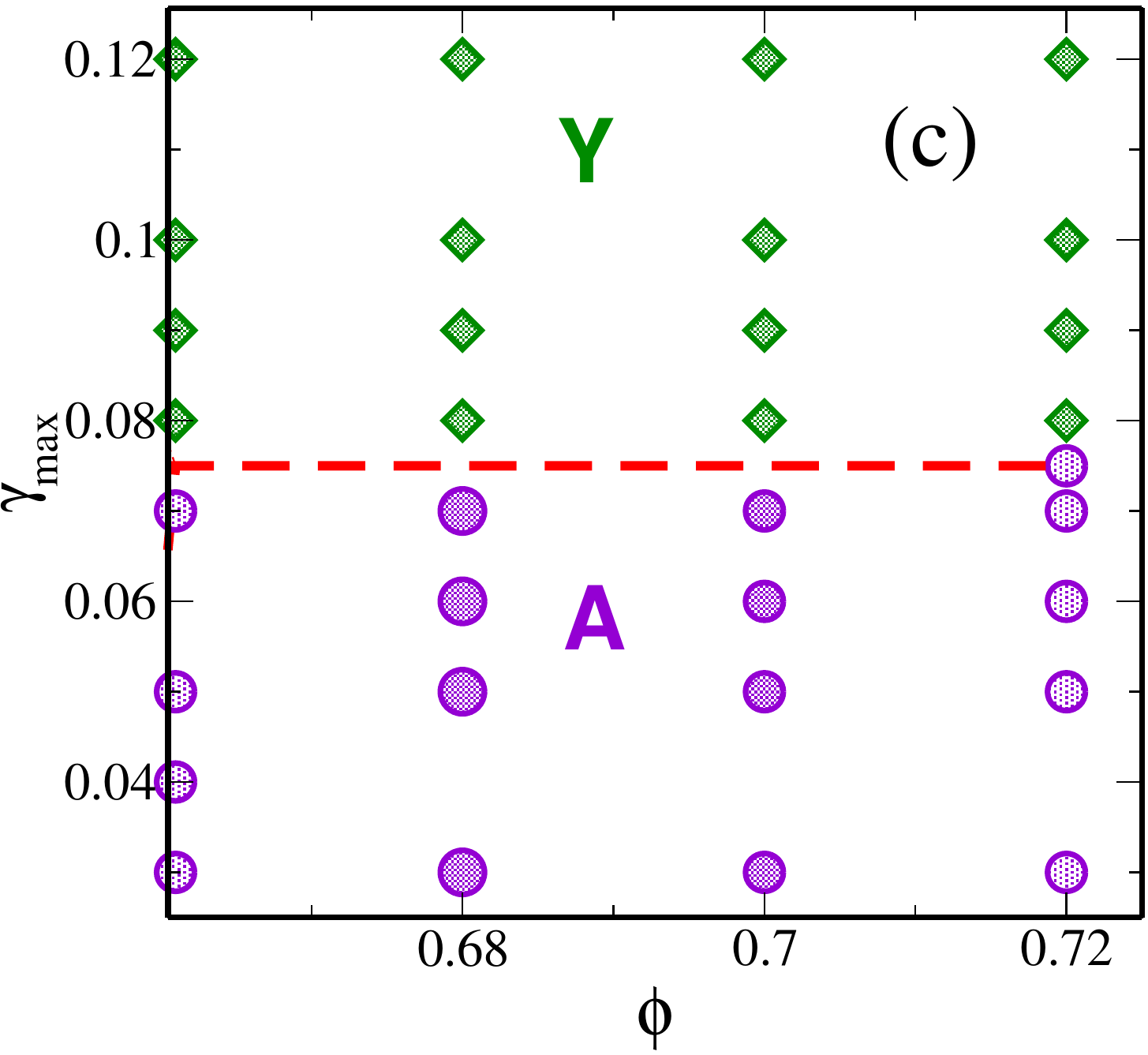}
\includegraphics[width=.25\linewidth]{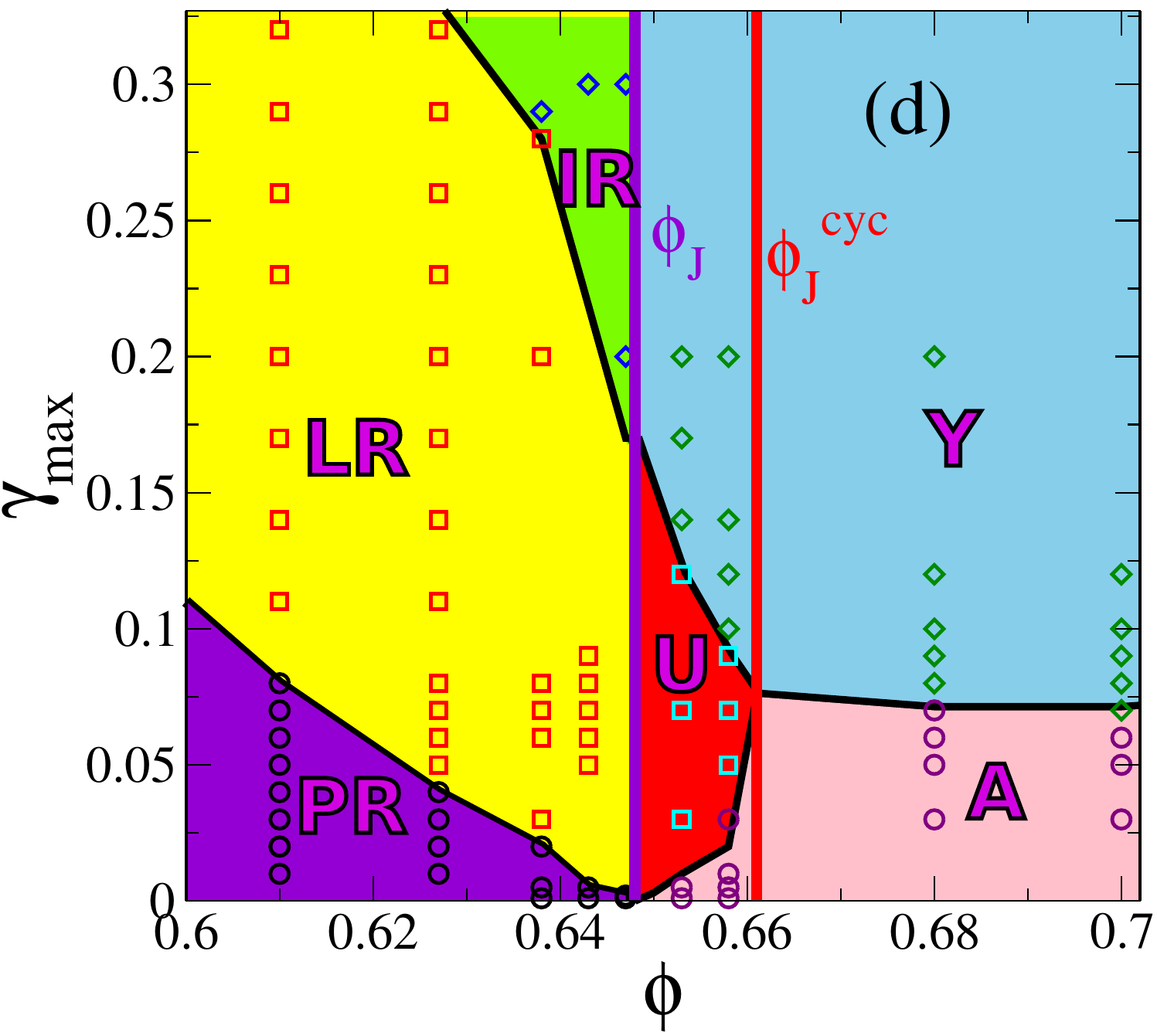}
\caption{The phase diagrams showing different transitions for different range of packing fractions ((a)-(c)) and the global phase diagram (d). Symbols in the legends indicate the following phases: PR: Point reversible; LR: Loop reversible; IR: Irreversible; Y: Yielded phase; U: Unjammed phase; A: Absorbing phase. {\bf (a)} Below $\phi_J$. \textbf{(b)} $\phi_J \geq \phi < \phi_J^{cyc}$. \textbf{(c)} $\phi > \phi_J^{cyc}$. (d) Complete phase diagram showing different phases and transitions across the isotropic jamming density.}
\label{phase}
\end{figure*}

Although the above analysis apparently offers a tidy grouping of all observed regimes into three types, one may ask if no further boundaries and transitions separate them further. There is indeed such a further separation, in the form of zero or finite energies and stresses. These quantities separate states below $\phi_{J}$ and those above, but with a curious exception. Moving from low to high densities at intermediate strain amplitudes at which the unjamming phase exists, energies do not become finite at $\phi_{J}$ but at higher densities that correspond to 
the transition from the unjamming phase to the absorbing phase, or the shear jamming phase. For strain amplitudes larger than the shear jamming value at $\phi_{J}$, the jump to finite energies occurs at $\phi_{J}$. Thus, the unjamming phase forms a curious zero energy and stress pocket in a finite energy regime for densities $\phi \geq \phi_{J}$.

The global phase diagram that emerges, shown in different density regimes in Fig. \ref{phase} (a) - (c) and in its entirety in \ref{phase} (d), has the point reversible phase at low densities and strain amplitudes, that terminates at the isotropic jamming density $\phi_{J}$. Starting with the loop reversible states below $\phi_{J}$ that lie at larger strains, moving to larger densities, one has a sequence of loop reversible states in the form of the unjammed phase, and the absorbing phase. At all densities, at still higher strain amplitudes, one sees a transition to the irreversible states. Interestingly, at all densities, the transition to the irreversible phase is characterized by a discontinuous jump in diffusion coefficients, which has in the past been shown to characterize the yielding transition in glasses under cyclic shear \cite{kawa2016,parmar2019strain}. The new remarkable observation is that all the transitions we analyse are associated with discontinuous changes in characterisations of trajectories (except the absorbing to unjammed states, characterised instead by a jump in contact number). The presence of loop reversible states at all densities is another general feature that is revealed by our results. Whether these special states are robust in the presence of thermal and other forms of noise remains to be investigated, but they are a common feature of athermally driven systems. Following recent work \cite{adhikari2018memory,mungan2019}, analyzing memory effects in the the different reversible regimes in light of the behaviour outlined in this work is of great interest. Although the unjamming phase is closely associated with the presence of a line of jamming points along the density axis, the isotropic jamming density (or minimum jamming density) $\phi_{J}$ emerges as a non-trivial threshold density. Point reversible states are confined to densities below this value, and 
energies and stresses become finite in the irreversible phase beyond this density. As noted however, energies and stresses remain zero above this density in the unjamming {\it pocket}, the origins of which merit further investigation.

\section{Summary and conclusions}
In summary, we have studied the reversible-irreversible transition below, close to and above the jamming density $\phi_J$. We have characterised different phases across the isotropic jamming density in detail by studying different microscopic quantities like the mean squared displacement, percentage of new collisions, non-affine path length, stress, potential energy, and contact numbers. For high density jammed packings, the reversible-irreversible transition corresponds to the yielding transition. We have demonstrated the presence of an unjamming region close to but above $\phi_J$. We identified $\phi_J^{cyc} = 0.661$ as the cyclic shear jamming density, above which the soft sphere packings behave like an elastic amorphous solid. Below $\phi_J$, two different forms of reversible phases are present, namely, point and loop reversible.  We showed that the non-affine path length and percentage of new collisions clearly distinguish reversible phases and irreversible phases for the whole range of densities. The transition to irreversible behavior is always characterised by the onset of diffusive behaviour of the particles across $\phi_J$. All transitions are characterised by discontinuous changes in relevant quantities.

Our work offers a comprehensive view of the response of particle assemblies to cyclic deformation, and is of relevance to a wide range of problems concerning the behaviour of driven amorphous particle assemblies. 
There are many other obvious directions in which our work can be extended, and we close by a brief discussion of some such directions. Despite much work, the origin of irreversible behaviour, in particular as a transition from periodic to chaotic states \cite{lavrentovich2017period,regev2017} remains incompletely elucidated. Our work has focused exclusively on frictionless sphere packings, but the role of friction in shear jamming is well appreciated and therefore understanding the implications of results for frictionless packings to the frictional case and extending the analysis here to the frictional case is of obvious importance. Some progress in that direction has been recently made \cite{otsuki2018shear,vinu2016,vinutha2019force}.  The other direction for further investigation is the role of thermal and other forms of noise on the behaviour revealed by our study. The observation of unjamming opens the exciting possibility of systematically studying (in progress \cite{varghese2019}) shear jamming and the related phenomenon of dilatancy in frictionless systems, and the possibility of a unified understanding of frictionless and frictional shear jamming.




\appendix
\section*{Appendix}

\subsection{Materials and Methods}

We consider a binary mixture (50:50) of $N=2000$ frictionless spheres interacting via a harmonic repulsive potential, $v_{ij} = \epsilon_{ij}(1-\frac{r_{ij}}{\sigma_{ij}})^2 $, where $\epsilon_{ij}$ is the interaction strength, and $\sigma_{ij} = (\sigma_i+\sigma_j)/2$, $\sigma_i$ the diameter of the $i^{th}$ particle. $\epsilon_{AA} = \epsilon_{BB} = \epsilon_{AB}$ and $\sigma_{B}/\sigma_{A} = 1.4$, and reduced units are defined in terms of $\epsilon_{AA}$ and $\sigma_{A}$.

We apply cyclic shear deformation using the athermal quasistatic protocol (AQS) which consists of an affine transformation to the coordinates ($x' = x+ d\gamma.z, y'=y, z'=z$), followed by energy minimization (conjugate gradient method (CG) except when specified otherwise), employing Lees-Edwards periodic boundary conditions. The stopping criterion for minimization is that the energy change between successive line minimizations, normalized to the energy value or the magnitude of the maximum component of force, falls below $10^{-16}$, whichever is satisfied earlier. For comparison with CG minimization, the FIRE \cite{bitzek2006structural} minimization method has been used (see Appendix, section 5). The system is subjected to cyclic deformation of amplitude $\gamma_{max}$ in small steps $d\gamma$, $= 10^{-3}$ and $=10^{-4}$ for densities below $\phi = 0.661$ and above respectively.

Initial configurations in the density range $0.54-0.627$ are obtained from hard sphere fluid configurations at $\phi=0.363$, subjected to fast compression using Monte Carlo simulations. The jamming density is estimated to be $\phi_J = 0.648$, following the method in \cite{pinaki-2010}.
Configurations close to $\phi_J$ ($0.638-0.647$) and above $\phi_J$ are obtained by a single step decompression or compression of configurations at $\phi_J$ followed by the energy minimization (CG).

The range of packing fraction for our study is $[0.54-0.72]$. The strain amplitude range is $[0.001-0.2]$ and $[0.01 - 1.0]$ for above and below $\phi_J$ respectively. For $\phi=0.56, 0.54$, $\gamma_{max}$ ranges from  $[0.1-8.0]$. To reach the steady state, we perform $\sim 10^2$ cycles in the irreversible regime and up to $10^3-10^{4}$ cycles in the reversible regime. Approximately 1950 independent simulations are used to construct the complete phase diagram. The number of independent samples used in different density ranges are - $10$ for  $0.661-0.72$, $10-20$ for $0.650-0.661$, $4-10$ for $0.638-0.648$, $6-10$ for $0.54-0.627$, and $1 - 2$ samples at the lowest densities at high amplitudes. 

\subsubsection{Definitions}
We compute the following quantities (along with PE and $\sigma_{xz}$) to characterize the steady states:\\
Mean squared displacement(MSD):  is defined as: 
\begin{equation} 
\triangle r^2(k) = \frac{1}{N}\sum_{i = 1}^N|r_i(k+k^o)-r_i(k^o)|^2 \end{equation} 
where $r_i(k)$ is the position at zero strain of $i^{th}$ particle in cycle number $k$. MSD is averaged over reference cycles $k^o$.\\ 
Average contact number ($Z$): is defined as
$Z = \frac{C}{N}$
where $C$ is the number of geometric contacts with $r_{ij} \leq \sigma_{ij}$.\\ 
Average non-rattler contact number ($Z_{NR}$): is defined as 
$Z_{NR} = \frac{C}{N-N_r}$
where $N_r$ is the number of rattlers in the system, {\it i.e.,} particles with less than $D+1$ contacts.\\
Average mechanical contact number ($Z_M$): is defined as: 
$Z_{M} = \frac{C_{M}}{N_{M}}$
where $C_{M}$ are the number of the mechanical contacts. A mechanical contact is identified when the overlap of a pair is greater than $(10^{-10})$(see Appendix, section $2$). $N_M$ is the number of particles that has finite force or stress. $Z_M$ can be $6$ and above or $0$, depending in a  jammed or unjammed phase respectively. Steady state quantities are indicated with a superscript $(SS)$.\\
Non-affine path length $(L)$: The non-affine path length is the average of the magnitude of non-affine displacement per particle within a cycle \cite{schreck2013particle}. 
\begin{equation} 
L^2(k) = N^{-1} \sum_i \{ \sum_n[(X^i_{k,n+1}-X^i_{k,n})^2+ (Y^i_{k,n+1}-Y^i_{k,n})^2 \nonumber 
\end{equation} 
\begin{equation}
+(Z^i_{k,n+1}-Z^i_{k,n})^2]^{\frac{1}{2}}\}^2
\end{equation}
where $k$ is cycle, $n$ is strain step, $X^{i}_{k,n}=x^{i}_{k,n} - \gamma_{n}z_{k,n} $, $Y^i_{k,n} = y^i_{k,n}$ and $Z^i_{k,n} = z^i_{k,n}$.\\ 
Percentage of new collisions ($C_{new}$): We compute the number of collisions that take place during a given cycle by number of pairs of particles that overlap ($r_{ij} < \sigma_{ij}$ when a stain step is applied. Considering a reference cycle $m$ in the steady state, we compute the number of collisions $C(m+k)$ at a later cycle $m+k$. The number of such collisions which also occurred in the $m$ cycle is defined as $C_m(m+k)$. The percentage of new collisions for the $k^{th}$ cycle (we typically use $k \geq 10$) after the reference cycle $m$ is thus:
\begin{equation}
C_{new}(k)=(1-\frac{C_{m}(m+k)}{C(m+k)})\times 100
\label{coleq}
\end{equation}

\subsection{Mechanical contacts}

\begin{figure}[ht!]
\centering
\includegraphics[width=.47\linewidth]{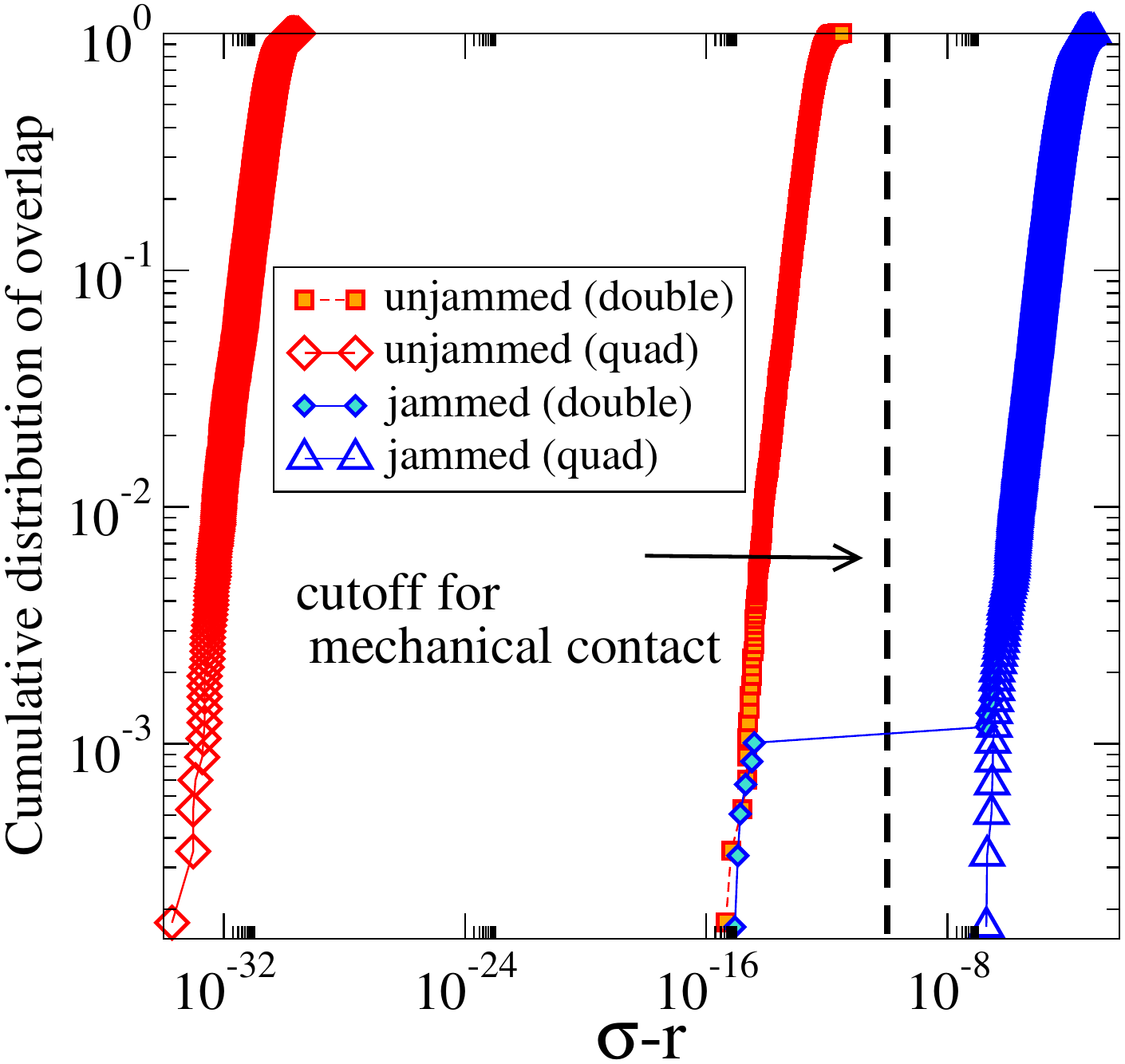}
\caption{Cumulative distribution of overlaps for a jammed and an unjammed packing at $\phi=0.653$, for two precision values. The vertical dashed line shows the tolerance value we use to identify $Z_M$.}
\label{mech_cut}
\end{figure}

\begin{figure}[ht!]
\centering
\includegraphics[width=.4\linewidth]{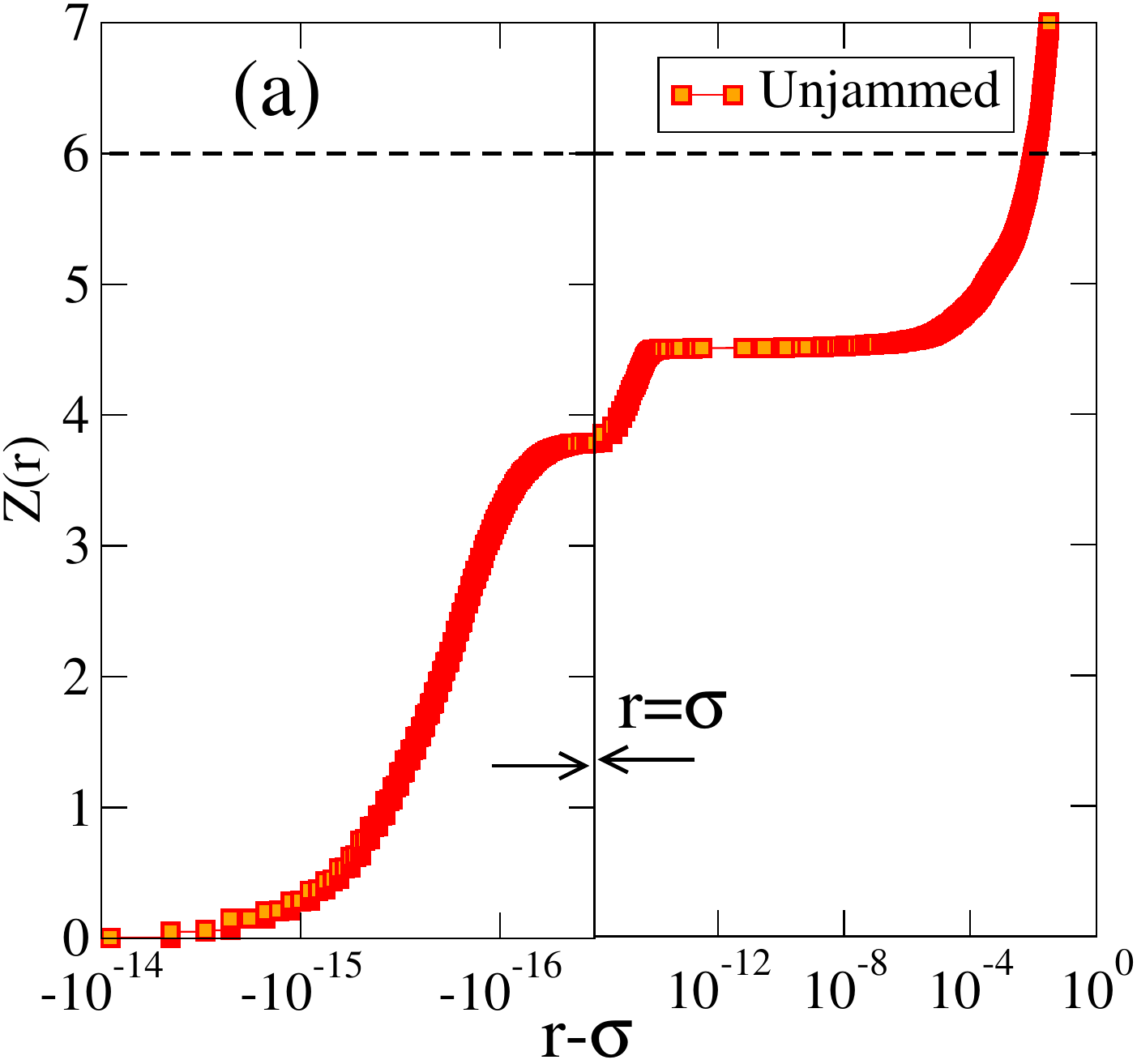}
\includegraphics[width=.4\linewidth]{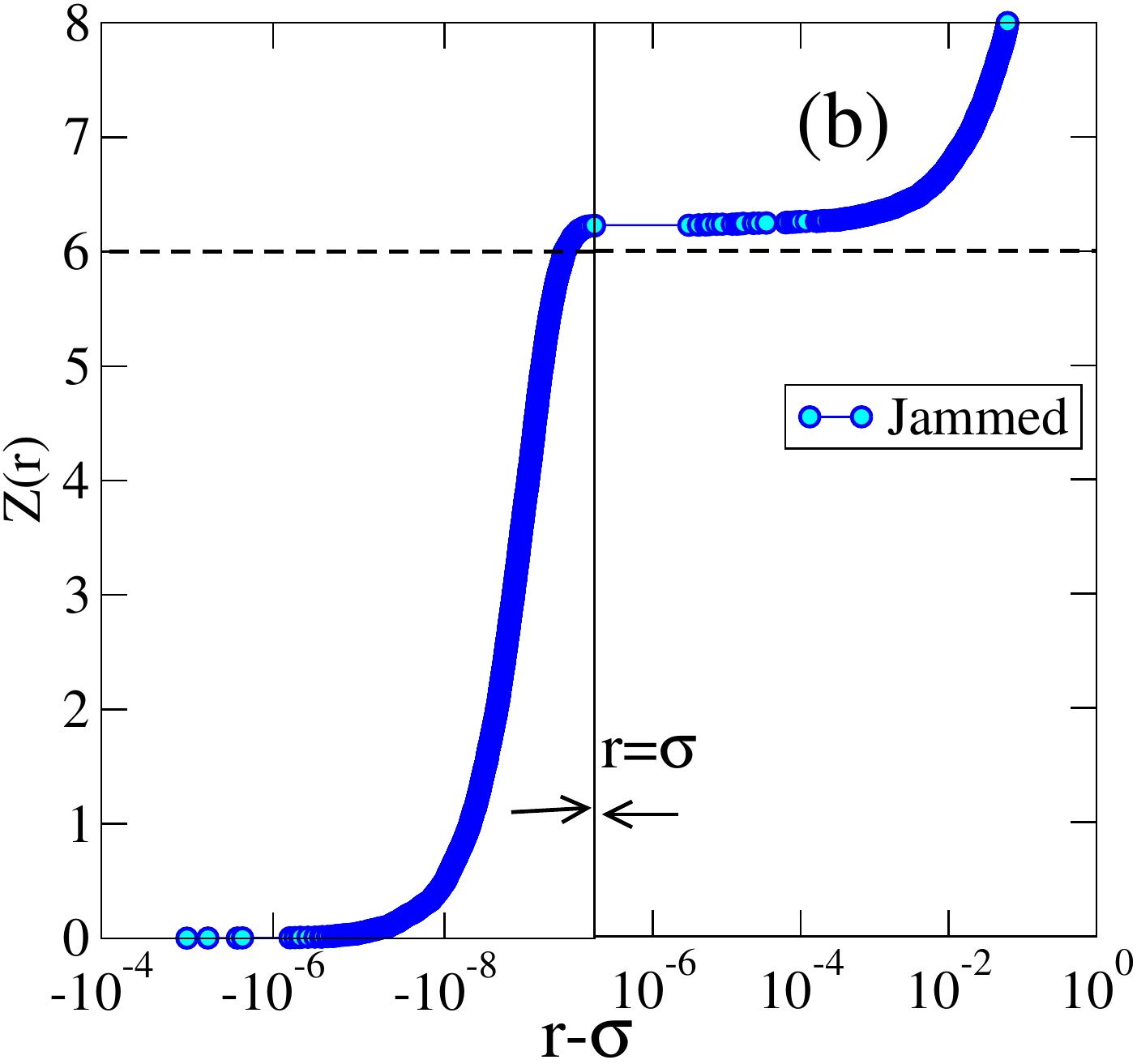}
\caption{The full cumulative distribution of contacts (interparticle separation $r$ both above and below $\sigma $) for an unjammed and a jammed configuration shows plateau below and above $2D$ respectively. The vertical lines indicate $r = \sigma$. }
\label{zofr}
\end{figure}

In jammed packings especially close to $\phi_J$ there are always rattler particles. Rattler particles do not belong to the stress carrying rigid network of a jammed packing. The percentage of rattler particles depends on the density of jammed packings and protocol. Higher the packing fraction lower is the percentage of rattlers. For frictionless particles, the geometric criterion for identifying rattlers is that they have contact number  $z<D+1$. A more robust criterion is a mechanical one, that is, rattlers are particles that do not carry any forces or stresses. However this mechanical criterion requires us to introduce a tolerance to identify a contact that carry stress due to the minimization protocol and the precision of computation. In Fig. \ref{mech_cut}, we show the cumulative distribution of overlaps for a jammed and an unjammed packing. Ideally the tolerance value is $r-\sigma = 0$ but due to the finite precision of the minimization and protocol details we use a tolerance value of $10^{-10}$ (shown as a vertical dashed line in Fig. \ref{mech_cut}). Observe that for a jammed packing, we can clearly see that the two distributions are well separated, {\it i.e.,} contacts that carry stress and those that do not. We also show the cumulative distribution for a quadruple (quad) precision minimization, see Fig. \ref{mech_cut}. We see that with quadruple precision, we can lower the tolerance value but the contacts that carry stress do not change. Note that in the jammed case, the contacts that have small overlaps with double precision computation, are pushed away when we perform quadruple precision computation. In Fig. \ref{zofr}, we show the full cumulative distribution of the contacts for a jammed and an unjammed configuration. For an unjammed configuration the plateau is below $2D$ and for a jammed configuration the plateau emerges above $2D$.

\subsection{Potential energy evolution near the yield strain}
\begin{figure}[ht!]
\centering
\includegraphics[width=.478\linewidth]{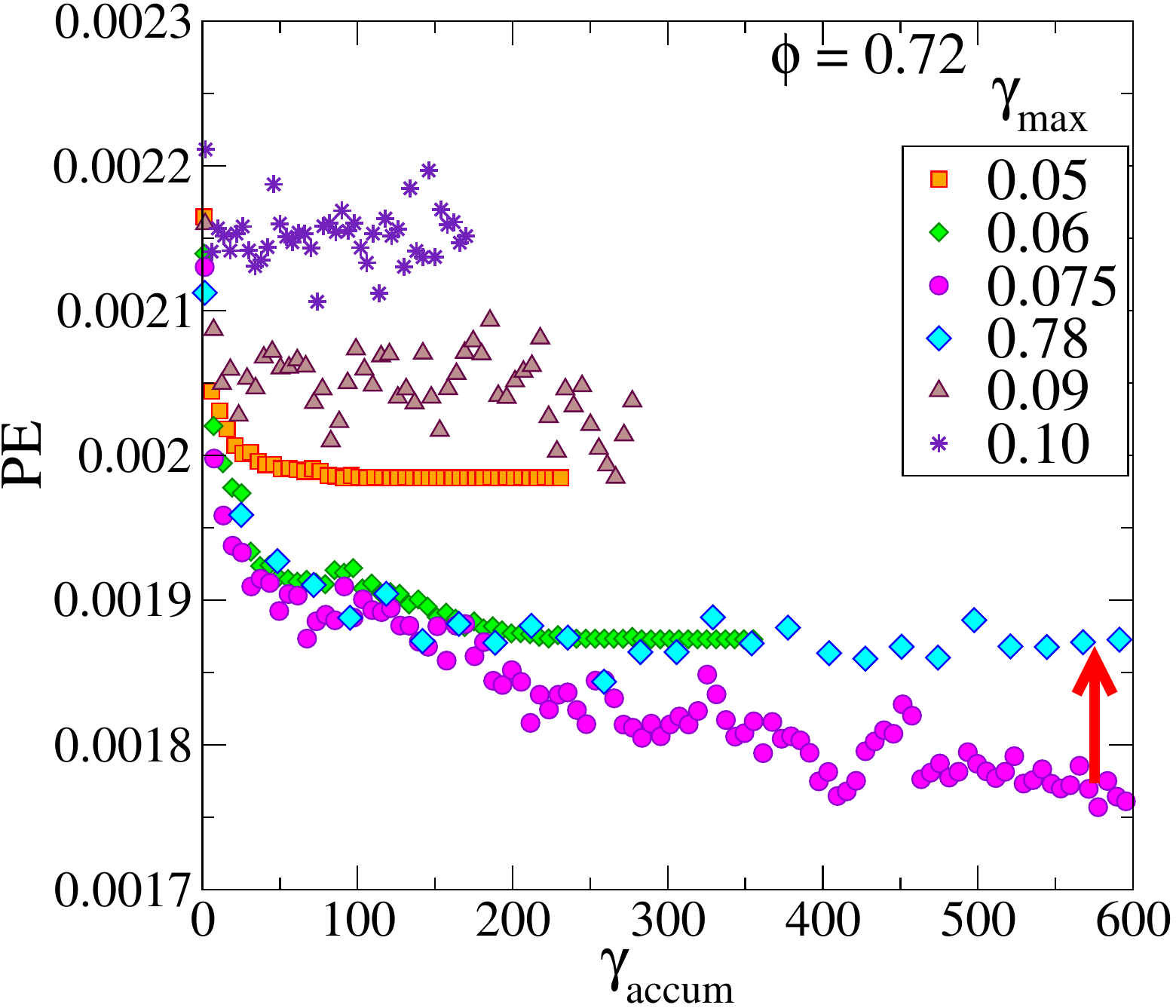}
\caption{ The potential energy PE attains a minimum value at the yielding transition amplitude which is identified as $\gamma_y = 0.075$ in our system. PE shows a discontinuous jump (red arrow) as the yielding point is crossed.} 
\label{pe072}
\end{figure}

We monitor the average potential energy (PE) as a function of number of cycles to test if the system has reached a steady state. In Fig. \ref{pe072}, we show the change in the potential energy (PE) of the system under different strain amplitudes. We observe that with the increase in strain amplitude $\gamma_{max}$, PE attains lower values for higher $\gamma_{max}$ in the steady state but above $\gamma_{max} = 0.075$, the steady state value of PE increases with $\gamma_{max}$. We also note that as we approach $\gamma_{max} = 0.075$, the yield strain amplitude, from above or below,  $\gamma_{accum}$ required to reach the steady state increases.  In the data shown, a steady state is reached in all the cases except $\gamma_{max} = 0.075$, at which the energies continue to evolve within the  $\gamma_{accum}$ window shown. 

\subsection{Contact number variation in the unjamming region}
\begin{figure*}[ht!]
\centering
\includegraphics[width=.3\linewidth]{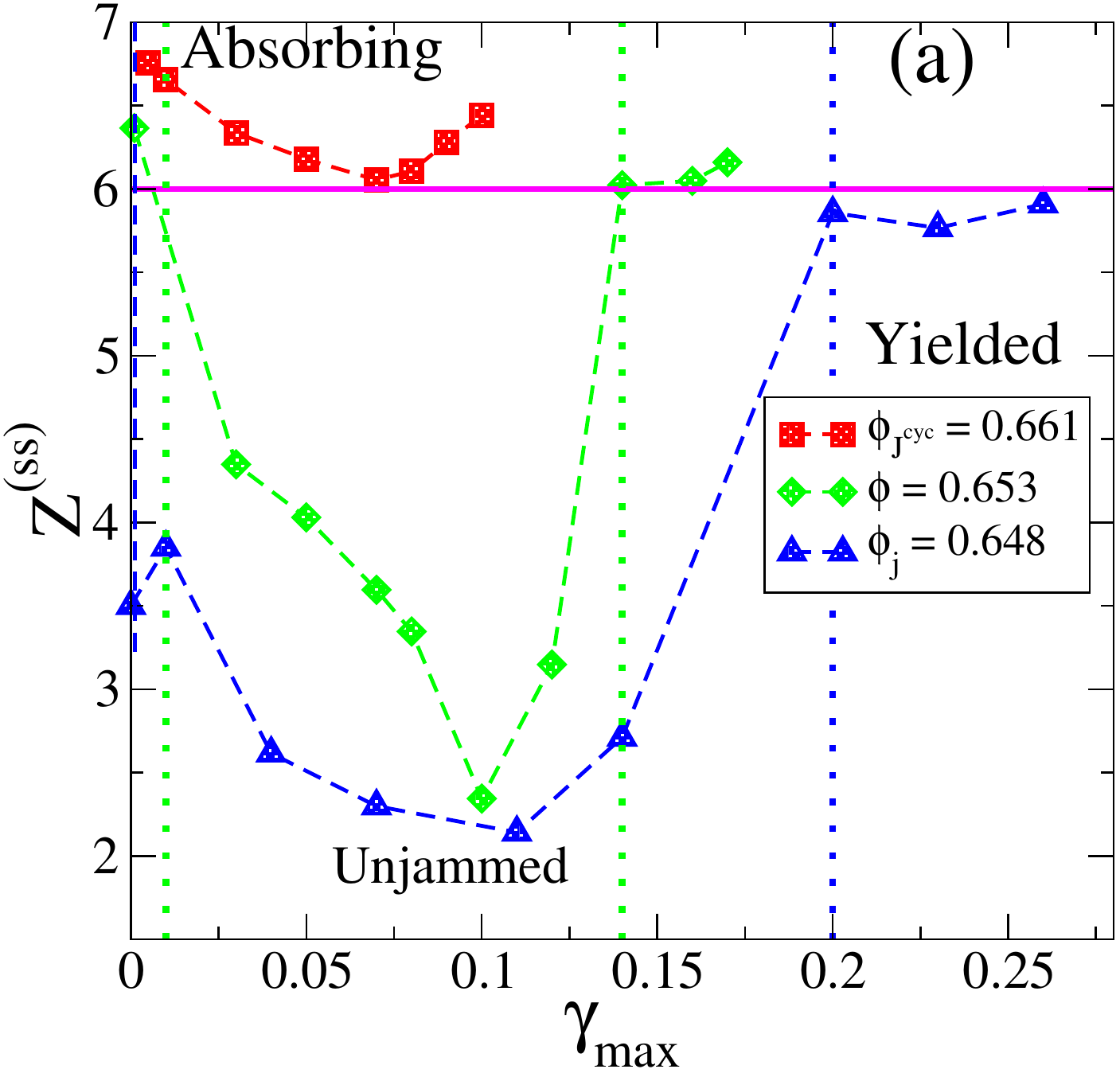}
\includegraphics[width=.32\linewidth]{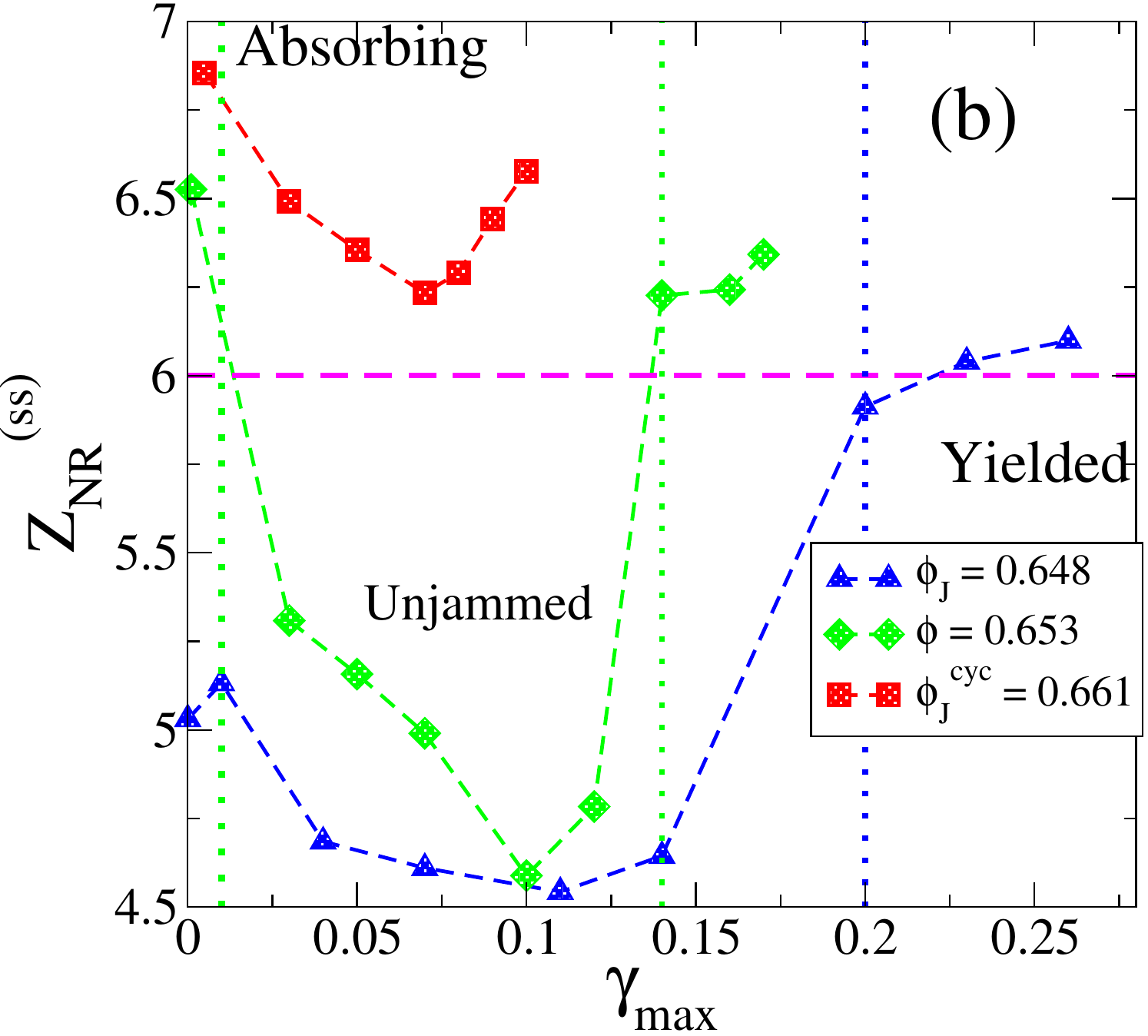}
\includegraphics[width=.3\linewidth]{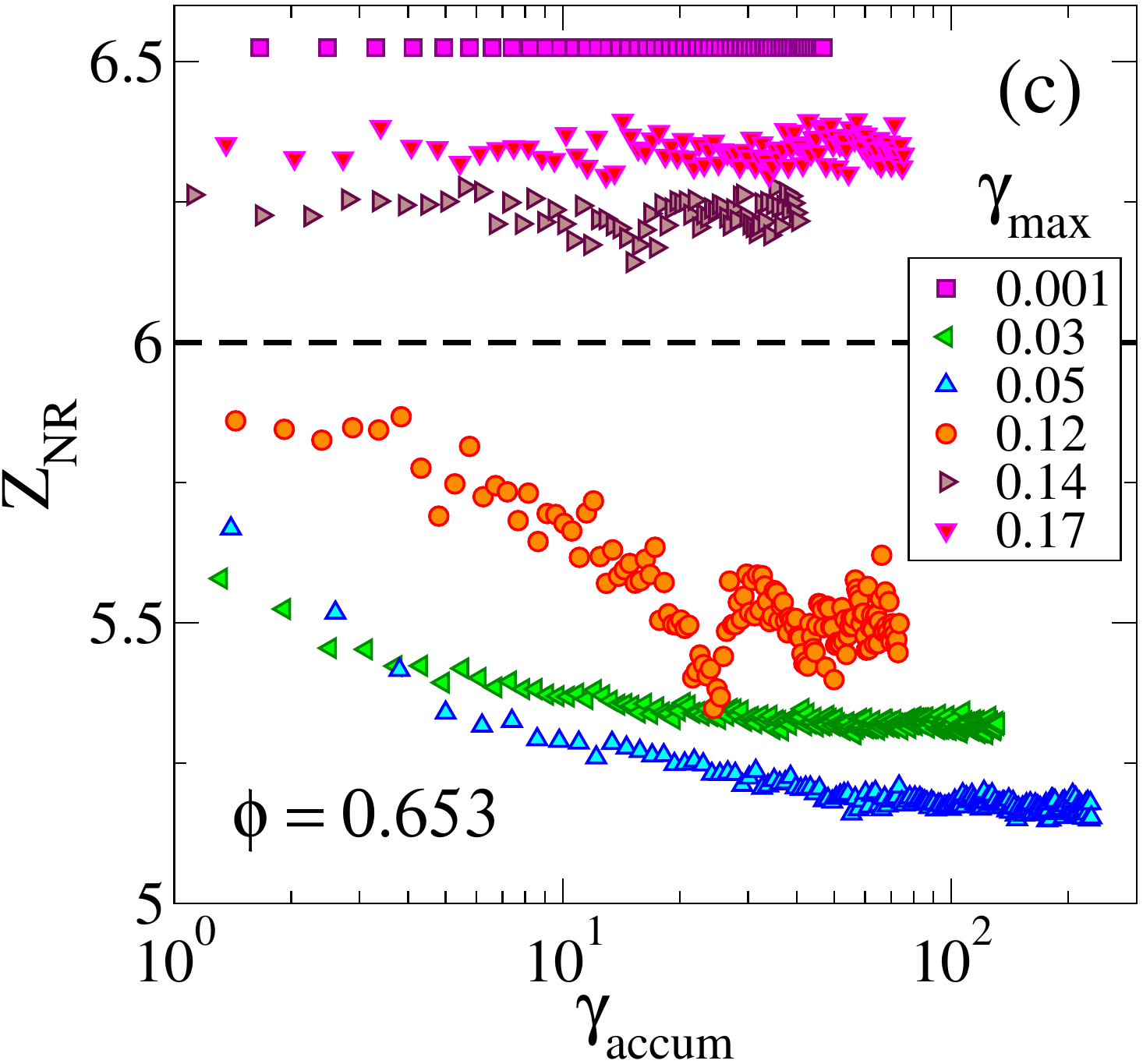}
\caption{The steady state average contact number $Z$ ({\bf (a)}) and contact number calculated without rattlers $Z_{NR}$ ({\bf (b)}) are shown as a function of strain amplitude, for $\phi_j \leq \phi < \phi_j^{cyc} = 0.661$.  $Z$ and $Z_{NR}$ drop to values below the isostatic contact value of $6$. The vertical line marks the boundaries of between the absorbing, unjamming and re-entrant jamming regimes. (c) The evolution of $Z_{NR}$ with $\gamma_{accum}$ for $\phi = 0.653$.} 
\label{zunjmall2} 
\end{figure*}

In Fig. $2$(a) of the paper, we show $Z_M$ as a function of $\gamma_{max}$, which shows that the system unjams for densities close to $\phi_J$. In Fig. \ref{zunjmall2} (a) and (b), we show the steady state value of the average contact numbers $Z$ and $Z_{NR}$, for different densities.
In Fig. \ref{zunjmall2}(c), we show the evolution of stroboscopic $Z_{NR}$ for $\phi=0.653$ and different strain amplitudes, which belongs to the unjamming region. We use this to identify steady state configurations. We observe that for $\gamma_{max}=0.001$ and $\gamma_{max} > 0.12$, the system in the jammed (finite stress) state and in the intermediate range of $\gamma_{max}$ the system is in the unjammed state. 

\subsection{Dependence of geometric contacts on the minimization protocol}
\begin{figure}[ht!]
\centering
\includegraphics[width=.47\linewidth]{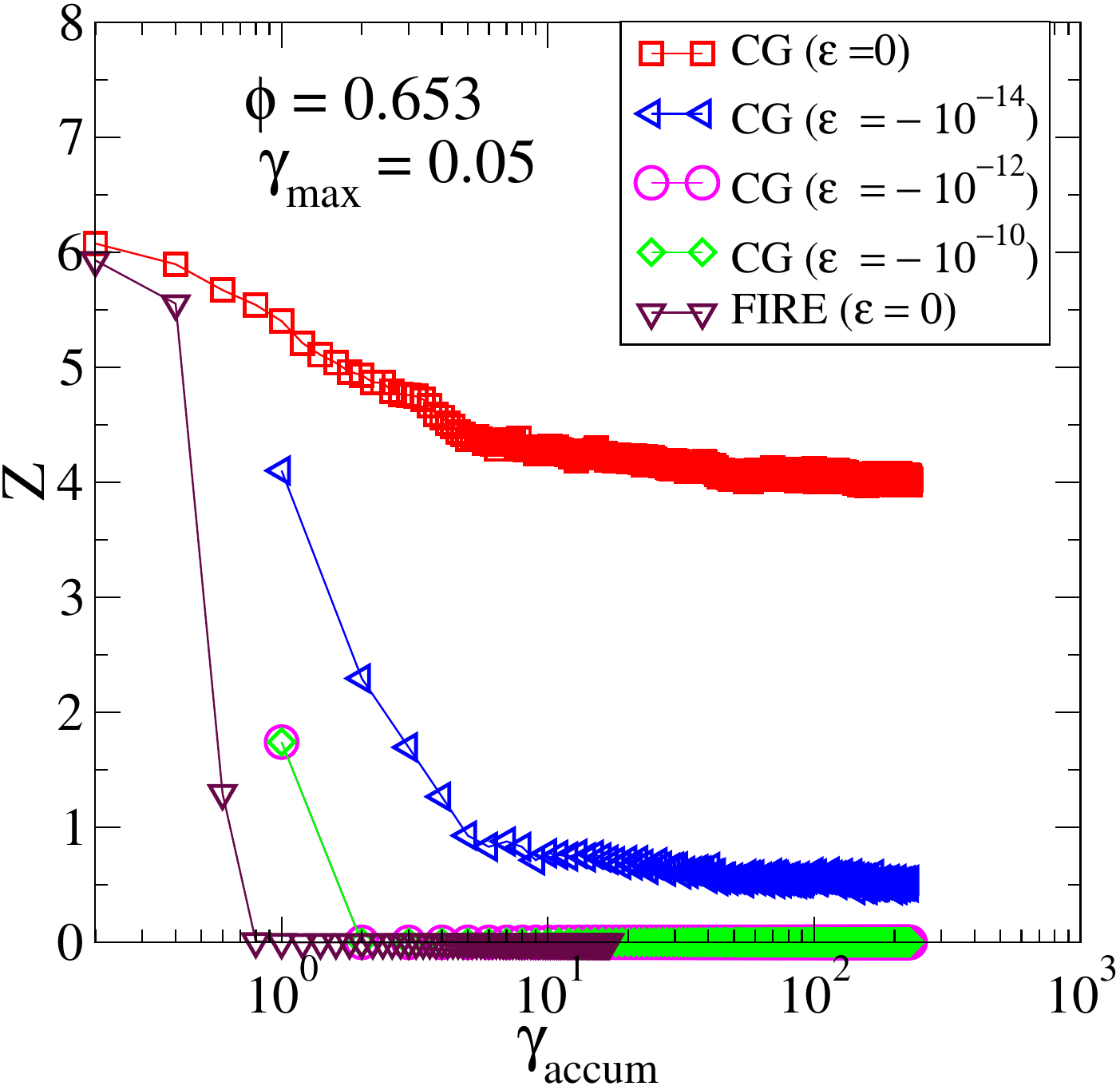}
\caption{The average contact number $Z$ as a function of $\gamma_{accum}$, shown for $\phi = 0.653$, for the CG method, with different tolerances, and the FIRE method. Compared to the conjugate gradient method, the Fire algorithm performs better minimization and removes all the contacts.}
\label{fire} 
\end{figure}

 The configurations in the unjamming region have a finite value of $Z$, see Fig. \ref{zunjmall2}(a). This is due to the minimization method used during AQS simulations. We use the FIRE minimization protocol during AQS steps and show that the unjammed configurations so obtained has $Z=0$. In Fig. \ref{fire}, we show the evolution of the average contact number as a function of $\gamma_{accum}$, for CG and FIRE protocol. We observe that the geometric contact number rapidly falls to zero when we use FIRE minimization. By using different tolerance $(\epsilon)$ values to identify the geometric contacts (like the mechanical contacts), $Z$ from the CG method approaches the $Z$ from the FIRE method, see Fig. \ref{fire}. Even though FIRE minimization method remove overlaps completely, computationally FIRE minimization method is almost one order of magnitude slower compared to CG minimization method.

\begin{figure}[ht!]
\centering
\includegraphics[width=.42\linewidth]{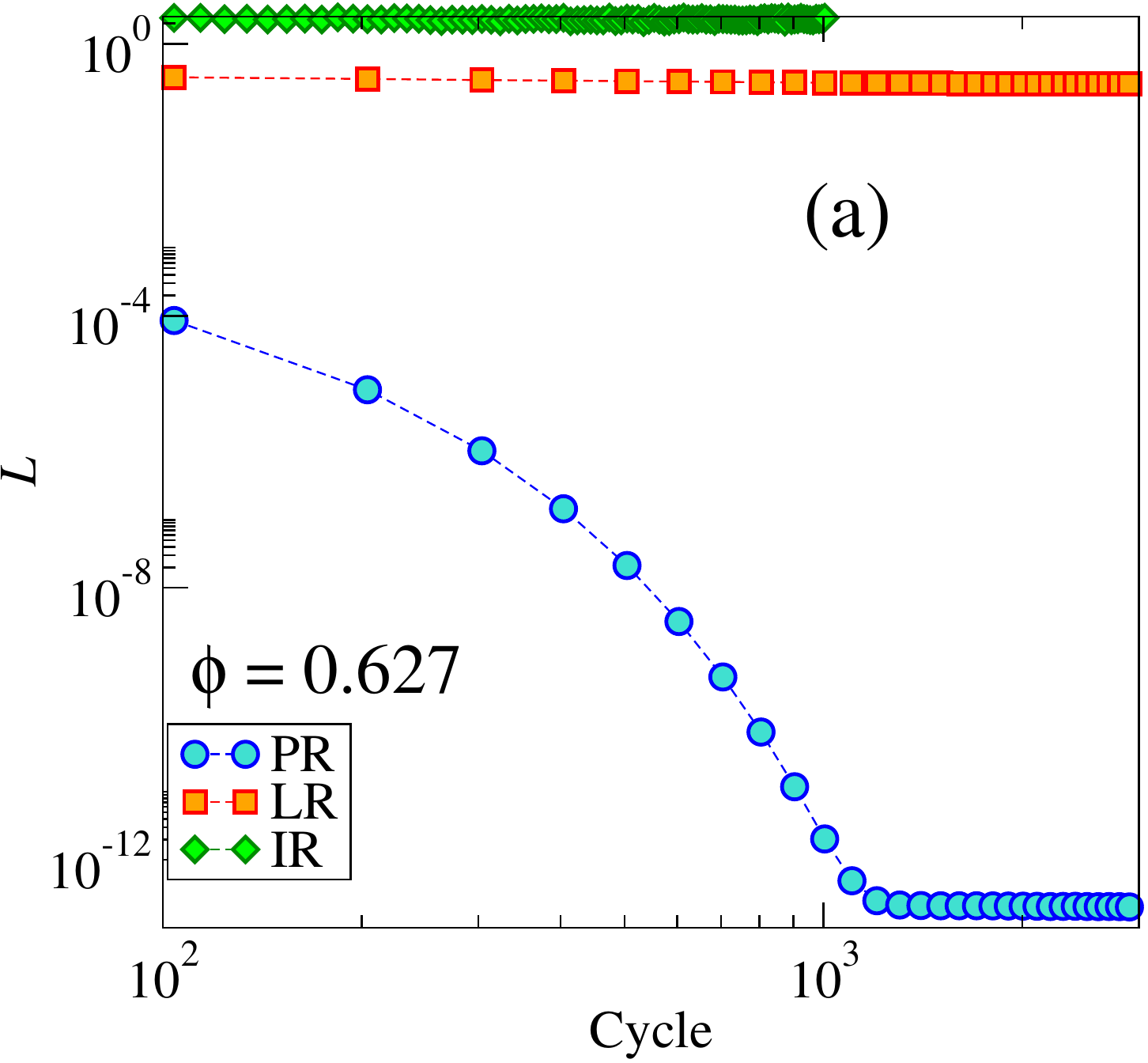}
\includegraphics[width=.42\linewidth]{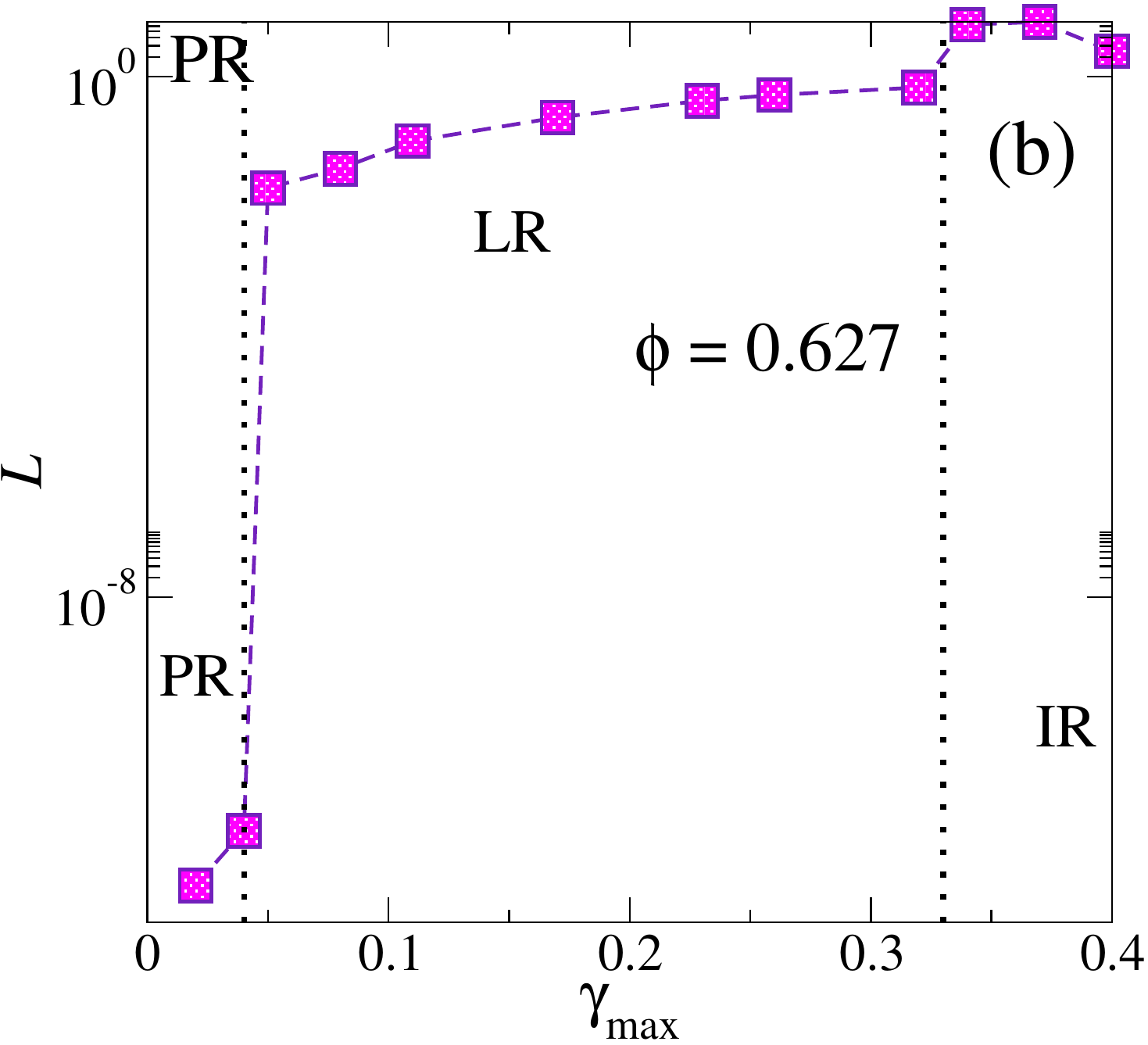}
\includegraphics[width=.42\linewidth]{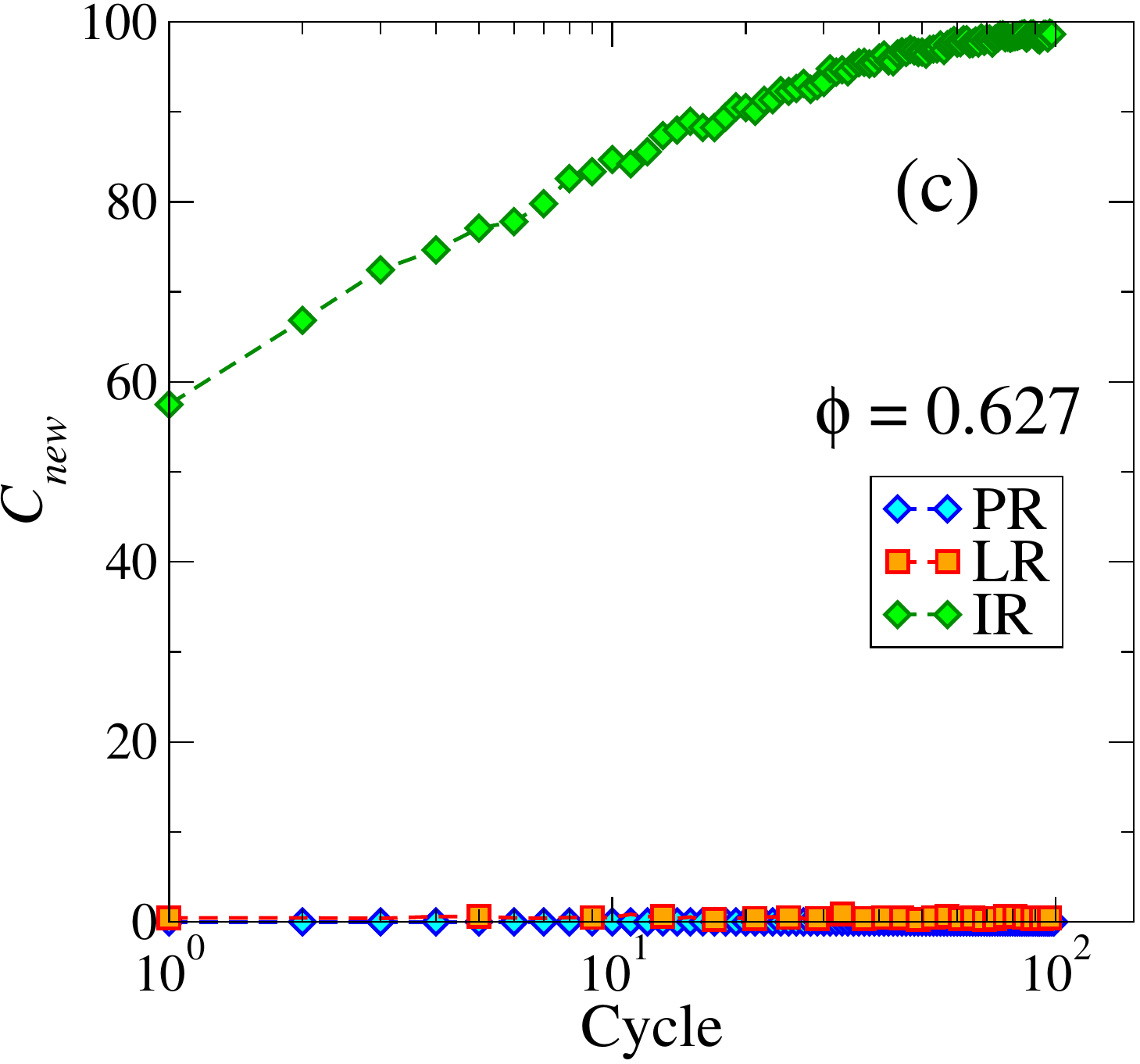}
\includegraphics[width=.42\linewidth]{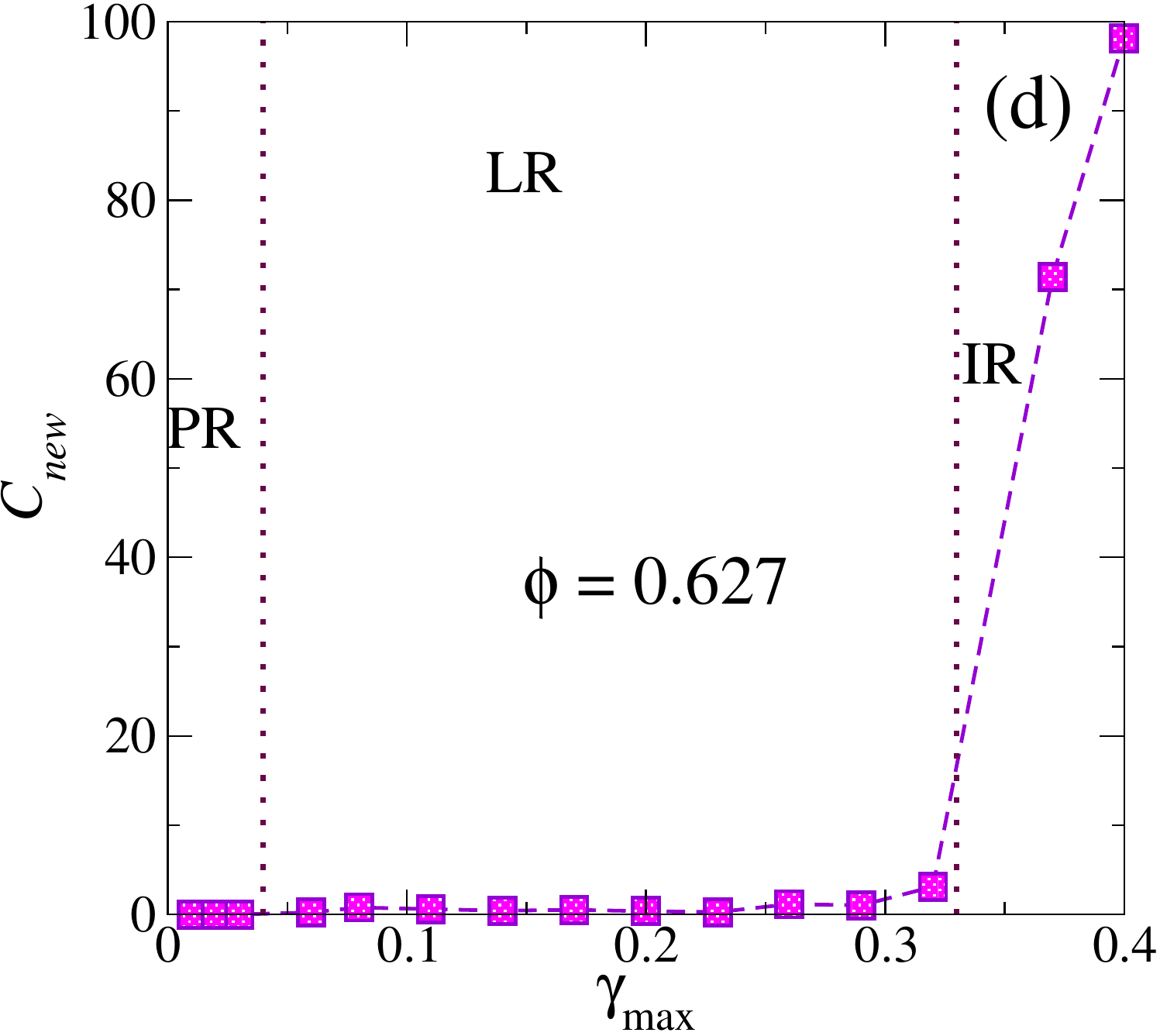}
\includegraphics[width=.42\linewidth]{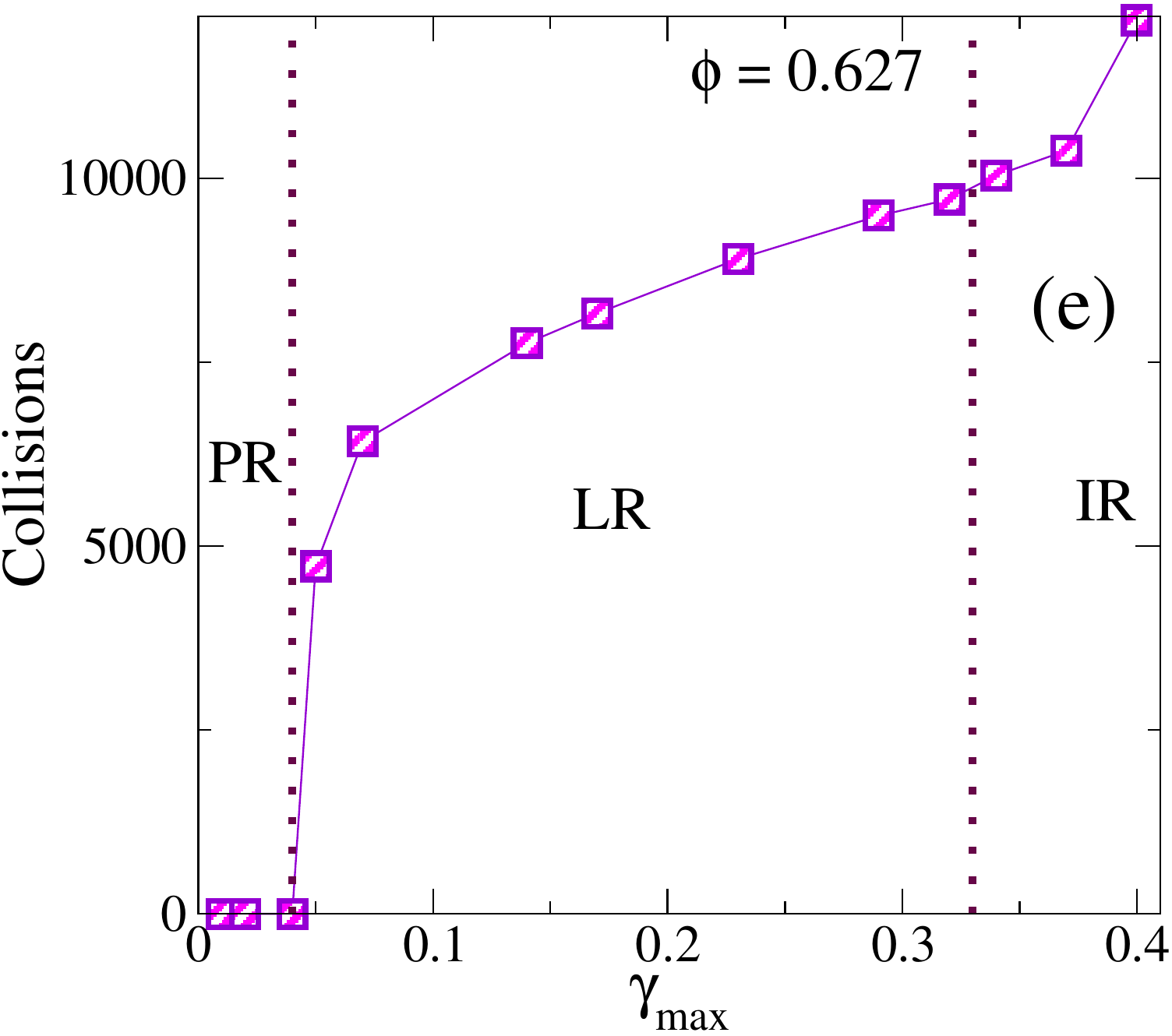}
\includegraphics[width=.42\linewidth]{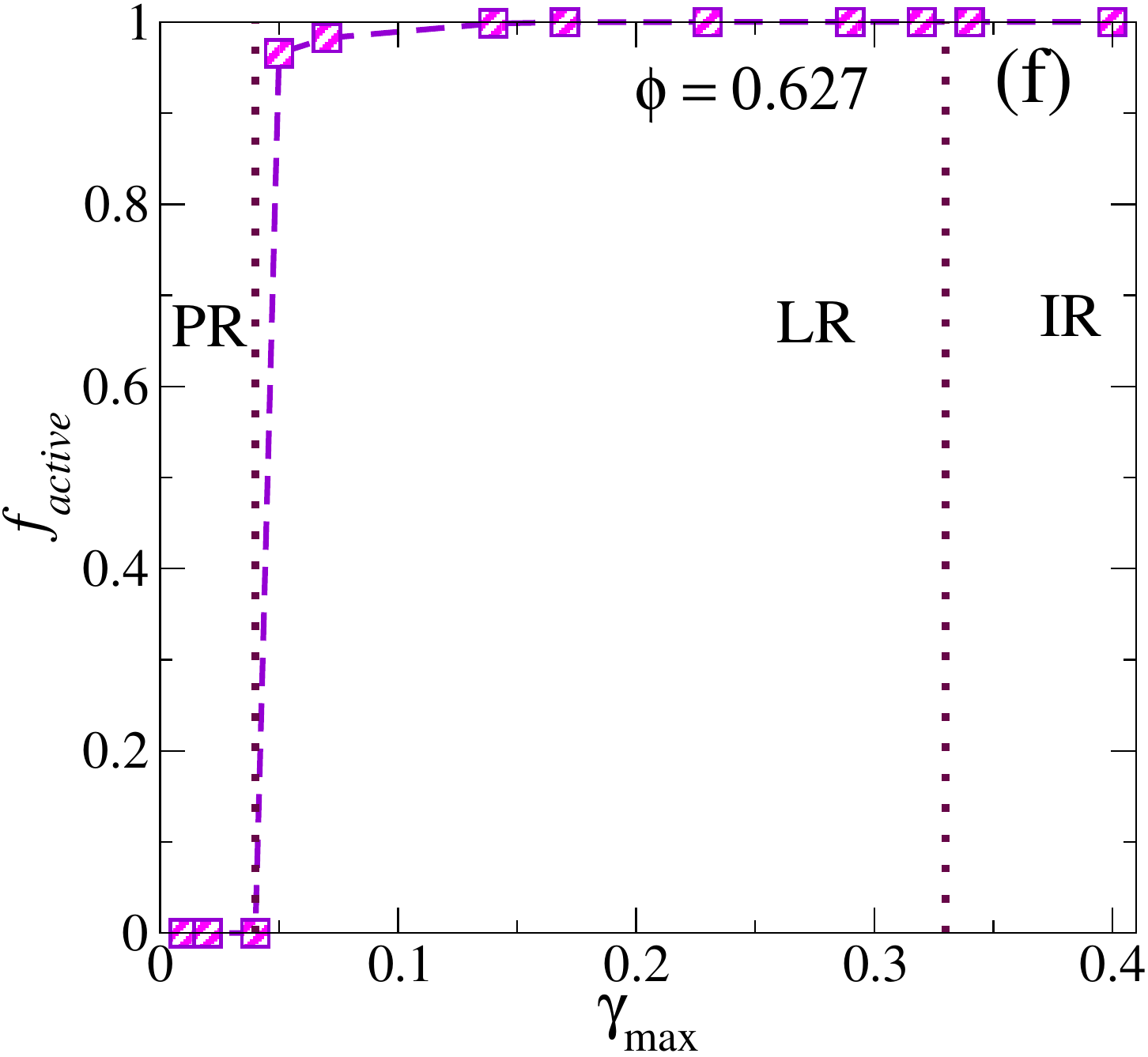}
\caption{{\bf (a)} Steady states are achieved as the non-affine path length reaches a steady state. {\bf (b)} The non-affine path length captures the transition from point reversible phase to loop reversible and irreversible states. {\bf (c)} $C_{new}$ as a function of $\gamma_{accum}$, shown for $\phi=0.627$ for all three phases. For PR and LR, $C_{new} = 0$ in the steady state and finite only for the IR state. {\bf (d)} Percentage of new collisions differentiates point and loop reversible states from the irreversible state. {\bf (e)} Total number of collisions, within a cycle, have been shown as a function strain amplitude, exhibiting a discontinuous jump across PR-LR boundary and shows a strain amplitude dependent increase across LR and IR phase. {\bf (f)} Fraction of active particles captures PR-LR transition through a discontinuous jump from $0$ (in the PR phase) to almost $1$ (in the LR phase).} 
\label{arcls}
\end{figure}

\subsection{Non-affine path length, collisions and active particles below $\phi_J$}
Because of the finite precision of the minimization, we find overlaps close to the machine precision even in the PR phase. In order to compute the collisions that take place, we, therefore, impose a cut-off of $10^{-10}$ for the overlaps arising between particles after the affine strain step but before minimization resulting in the final configuration. We compute the non-affine path length, the percentage of new collisions ($C_{new}$) and fraction of active particles to characterise different phases and to identify the transitions. For the point reversible states (PR), particles trace the same path in a cycle (as there are no collisions ) and hence $L$ and $C_{new}$ is zero. For the loop reversible states (LR), collisions occur in a cycle, and hence the non-affine path length is finite. Also, in the loop reversible states, even though collisions occur, they occur between the same pairs and hence $C_{new} = 0$, see Fig. \ref{arcls}. Only in the diffusive region, irreversible states (IR), $C_{new}$ is finite. In Fig. \ref{arcls}(a), we show $L$ as a function of $\gamma_{accum}$, which we have used to identify the steady state configurations. When a particle at least collides once within a cycle during the affine transformation, in its steady state, we identify such a particle as an active particle. Even though the number of collisions shows a strain dependent increase (see Fig. \ref{arcls}(e)) beyond PR phase, almost all the particles seem to collide at least once in a cycle in LR or IR phase (Fig. \ref{arcls} (f)). To summarise,  PR to LR transition is identified by the discontinuous jump in $L$, and fraction of active particles whereas LR to IR transition is captured through the discontinuous increase of $C_{new}$ and $L$.

\subsection{Timescales across the phases below $\phi_J$}

\begin{figure}[ht!]
\centering
\includegraphics[width=.46\linewidth]{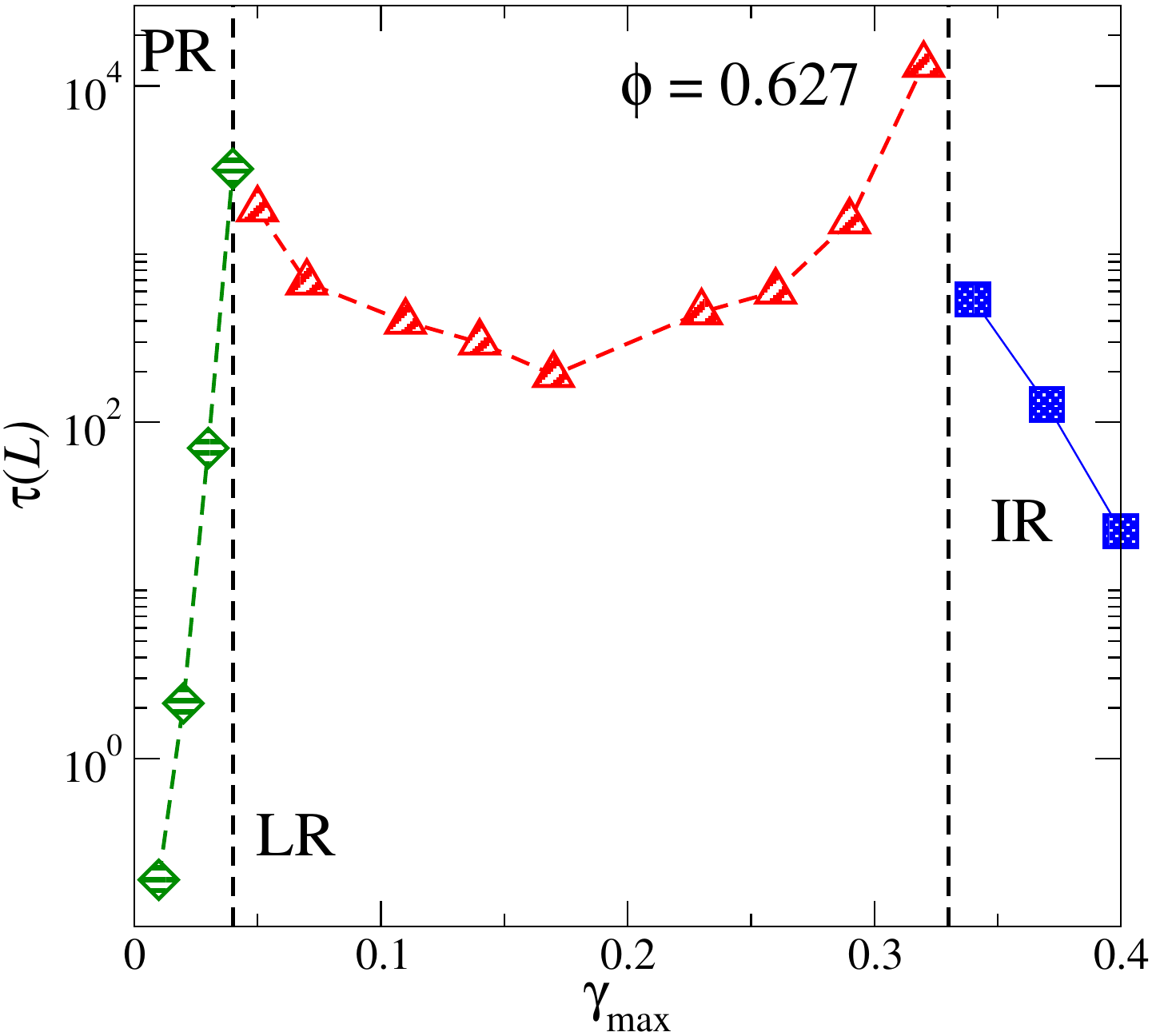}
\caption{The time scale extracted from the relaxation of $L$ shows a non-monotonic change across the PR-LR and LR-IR boundaries. The timescale ($\tau$) has been extracted from the exponential relaxation in PR phase and a stretched exponential relaxation in the LR and IR phase of the non-affine path length.}
\label{tscale} 
\end{figure}

The non-affine path length $L$ that we have used to identify and characterise different phases below $\phi_J$, captures the slow down of relaxation to the steady state near the PR - LR and LR - IR transitions. The timescale ($\tau$) has been extracted from the relaxation of the $L$ as the system approaches a steady state with increasing $\gamma_{accum}$. The point reversible phase shows an exponential decay $(L(\gamma_{accum})=L(0)e^{-(\gamma_{accum}/\tau)})$, whereas the loop reversible phase and the irreversible phase show a stretched exponential decay ($L(\gamma_{accum})=(L(0)-L(\infty))e^{-({\gamma_{accum}}/{\tau})^{\beta}}+L(\infty)$) of the non-affine path length. We see an initial increase of the $\tau$ as we approach  PR-LR boundary (see Fig \ref{tscale}) and we see a similar increase in timescale across LR-IR boundary after an intermediate decrease of $\tau$ in the middle region of the LR phase. This behaviour indicates the existence of two different transitions at the upper and lower limit of the loop reversible phase. The results shown do not permit us to conclude that the time scales diverge at the transitions, and a more careful analysis close to the transitions is required to make any definite statements in this regard. 

\bibliography{softsphere}

\end{document}